\documentclass[a4paper,pre,superscriptaddress,floatfix,nofootinbib,twocolumn]{revtex4-2}

\usepackage{bbold}
\usepackage{bbm}
\usepackage[pdftex]{graphicx}
\usepackage{latexsym,amsmath,verbatim,amssymb,txfonts, mathtools}
\usepackage{color}
\usepackage{rotating}
\usepackage{verbatim}
\usepackage{multirow}
\usepackage[english]{babel}
\usepackage{comment}
\usepackage{hyperref}
\usepackage{mathrsfs}

\DeclarePairedDelimiter\floor{\lfloor}{\rfloor}

\newcommand{\av}[1]{\langle {#1} \rangle}

\begin{document}


\title{Critical avalanches of Susceptible-Infected-Susceptible dynamics in finite networks}

\author{Daniele Notarmuzi}
\affiliation{Institut für Theoretische Physik, TU Wien, Wiedner Hauptstraße 8-10, A-1040 Wien, Austria}
\affiliation{Center for Complex Networks and Systems Research, Luddy School
  of Informatics, Computing, and Engineering, Indiana University, Bloomington,
  Indiana 47408, USA}
\email{daniele.notarmuzi@tuwien.ac.at}

\author{Alessandro Flammini}
\affiliation{Center for Complex Networks and Systems Research, Luddy School
  of Informatics, Computing, and Engineering, Indiana University, Bloomington,
  Indiana 47408, USA}

\author{Claudio Castellano}
\affiliation{Istituto dei Sistemi Complessi (ISC-CNR), Via dei Taurini 19, I-00185 Rome, Italy}
\affiliation{Centro Ricerche Enrico Fermi, Piazza del Viminale 1, 00184 Rome, Italy}

  \author{Filippo Radicchi}
\affiliation{Center for Complex Networks and Systems Research, Luddy School
  of Informatics, Computing, and Engineering, Indiana University, Bloomington,
  Indiana 47408, USA}

\begin{abstract}
We investigate the avalanche temporal statistics of the Susceptible-Infected-Susceptible (SIS)
model when the dynamics is critical and takes place on finite random networks.
By considering numerical simulations on annealed topologies we show that the
survival probability always exhibits three distinct dynamical regimes.
Size-dependent crossover timescales separating them scale differently for
homogeneous and for heterogeneous networks.
The phenomenology can be qualitatively understood based on known features of the
SIS dynamics on networks.
A fully quantitative approach based on Langevin theory is shown to perfectly
reproduce the results for homogeneous networks, while failing in the heterogeneous case.
The analysis is extended to quenched random networks, which behave in agreement
with the annealed case for strongly homogeneous and strongly heterogeneous networks.
\end{abstract}

\maketitle

\section{Introduction}

Systems undergoing continuous absorbing-state phase-transitions are, exactly at the critical point, infinitely susceptible
to local perturbations.
This implies that, introducing a single active seed in the absorbing state,
the subsequent evolution leads to an avalanche of activation events that may span
the whole range of temporal and spatial scales.
Examples of this avalanche dynamics abound, both in the realm of physics and in other
biological, social and technical
domains~\cite{beggs2003neuronal,wang2013self,bak2002unified,kinney2005modeling,
nishi2016reply,wegrzycki2017cascade,lerman2010information}.

The multiscale nature of these phenomena is quantitatively characterized by
the size and duration distributions of avalanches.
If the control parameter is subcritical (i.e., in the absorbing phase) size and duration of avalanches are
exponentially distributed,
extending up to well defined and finite temporal and spatial scales.
In the supercritical domain, instead, a finite fraction of them spans the whole system, leading
(in infinite systems) to a stationary active state.
In the critical case all avalanches end in a finite time, but the distributions have power-law tails so that
events of any size and duration are possible.
Denoting with $z$ the size of an avalanche and with $t$ its duration, we can in general write
for the probability distributions of size $z$ and duration
$t$~\cite{Kadanoff1989}
\begin{equation}
~~~~
P(z) \sim z^{-\tau} {\cal F} (z/z_{\times})
\; \textrm{ and } \; 
P(t) \sim t^{-\alpha} {\cal G} (t/t_{\times}),
\end{equation}
where $\tau$ and $\alpha$ are universal critical exponents, 
${\cal F}$ and ${\cal G}$
are scaling
functions and the cutoff scales 
$z_{\times}$ 
and $t_{\times}$ 
depend at criticality only on the system size.
Averaging the avalanche size at a fixed duration $t$ it is possible to define another
exponent 
$\langle z \rangle \sim t^{\theta}$, 
which is related to the others
as~\cite{sethna2001crackling}
\begin{equation}
\theta = \frac{\alpha-1}{\tau-1}.
\end{equation}

The branching process (BP) model~\cite{harris1963theory} provides the natural
framework for describing (at least approximately) a large class of systems undergoing continuum absorbing
phase transitions. 
In the simplest case, when the distribution of offsprings (i.e., the probability
for an active element to activate a given number of new elements) has finite variance,
BP theory predicts
\begin{equation}
\tau = 3/2 , 
~~~~
\alpha = 2, 
~~~~
\; \textrm{ and } \; 
\theta = 2.
\end{equation}

This kind of mean-field behavior is expected to occur when the system dynamics
takes place on a homogeneous network,
where the second moment of the degree distribution is finite.
The phenomenology changes when the distribution of offsprings in the 
BP
has infinite variance, as it happens for processes taking place
on heterogeneous networks with degree distribution $P(k) \sim k^{-\gamma}$
and $2<\gamma \leq 3$.
In such a case BP theory predicts anomalous, $\gamma$-dependent,
exponents~\cite{Adami2002, Goh2003, Saichev2005, gleeson2014competition}
\begin{equation}
\tau = \frac{\gamma}{\gamma-1},  ~~~ \alpha=\frac{\gamma-1}{\gamma-2}, 
~~~
\; \textrm{ and } \; 
\theta = \frac{\gamma-1}{\gamma-2}.
\label{eq:expo_anomal}
\end{equation}

While standard BP behavior is observed for many systems on homogeneous
networks~\cite{di2017simple,Zapperi1995,gleeson2017temporal,matin2021scaling},
recent work has cast doubts over the applicability of this theoretical
framework to many avalanche phenomena in heterogeneous networks~\cite{radicchi2020classes}.
In all cases considered, apart from a possible preasymptotic regime
valid for short duration and small size, the distributions decay with
exponents in agreement with standard BP values, with no dependence on $\gamma$.
Possible causes of this violation of the BP predictions have been put forward
in Ref.~\cite{radicchi2020classes}. One possible reason for such
violation is the existence of loops in networks, possibly allowing for 
nodes that are already active to be reached again by the infection, which is 
an explicit violation of the mapping between spreading models and the BP.
Further, finite networks always have a 
finite second moment of the degree distribution, which prevents the anomalous 
exponents~(\ref{eq:expo_anomal}) to be asymptotically observed even in a pure
BP. It was also noted in Ref.~\cite{radicchi2020classes} that the spreading 
mechanism of certain models, such as the contact process (CP), do not really
involve all the neighbors of a node and hence avalanches are not impacted by
unbounded fluctuations of the network degree distribution. 
These hypotheses,
however, 
call
for more detailed investigations of the origin of the
breakdown of anomalous BP exponents in specific systems
and for the formulation of alternative theoretical
approaches such as the one presented in Ref.~\cite{larremore2012statistical}.
Also, the way avalanches depend on the system size in finite systems, which
is an issue of crucial importance for the analysis of real systems,
is still largely unexplored.

In this work we consider Susceptible-Infected-Susceptible (SIS)
dynamics~\cite{PastorSatorras2015},
one of the most fundamental (and simple) models deemed to be described by BP,
and perform such an investigation in complex networks of variable size.
Apart from the intrinsic importance of SIS dynamics,
an additional motivation for our study is that it has been
realized that the interplay between heterogeneous topology and SIS dynamics
gives rise to highly nontrivial phenomena, such as the vanishing of the
threshold in the large-network
limit~\cite{Pastor2001epidemic,castellano2010thresholds,Boguna2013},
the interplay between distinct subextensive subgraphs~\cite{castellano2012competing}
and long-range percolation effects~\cite{Castellano2020}.
Whether and how these nontrivial properties of the stationary state are reflected
also in 
the avalanche statistics
and its dependence on the system size 
are other aspects
deserving to be analyzed.

Our investigation is inspired by the work of Ref.~\cite{boguna2009langevin} that deals
with critical properties of the 
CP, 
a model akin to SIS,
but exhibiting simpler critical dynamics characterized by a finite threshold
for any value of $\gamma$.
For CP, avalanche exponents are nonanomalous also for $2 < \gamma \leq 3$,
but this is easily reconciled with the BP phenomenology,
as the effective distribution of the number of offsprings does not depend
on the network substrate and is always homogeneous.
Yet, Ref.~\cite{boguna2009langevin} presents a detailed analytical investigation of CP
based on a Langevin approach, which constitutes the natural basis also
for the analytical approach to SIS avalanches and in particular 
for their temporal
properties.




In this paper we perform numerical simulations of the critical SIS model on annealed and quenched networks, both homogeneous and heterogeneous, generated according to the configuration model (CM)~\cite{molloy1995critical}.
 We further develop a theoretical approach based on the use of Langevin
equations to study the probability that avalanches have duration at least $t$. The theoretical approach
explicitly makes use of the annealed network approximation, but we show that it further
provides with a partial understanding of the
avalanche behavior 
on quenched networks.
The rest of the paper is organized as follows.
In Section~\ref{sec:model}, the dynamics of the SIS
model and the statistical features of the networks we consider are briefly introduced. In 
Section~\ref{sec:summary}, we summarize the phenomenology we observe on different network structure, and offer a clear physical 
interpretation
of the avalanche behavior. In Section~\ref{sec:annealed_results}, numerical results for annealed networks are discussed. In Section~\ref{sec:langevin_main}, the theoretical
approach is presented and its predictions are compared with the results of the previous section.
In Section~\ref{sec:quench_results},
results for quenched networks are shown and compared to the theoretical predictions of the previous section. We discuss our results Section~\ref{sec:conclusions}.


\section{Model definitions}
\label{sec:model}

\subsection*{Spreading dynamics}

We consider the continuous-time Susceptible-Infected-Susceptible (SIS) dynamics
on networks.  Each node can be either infected (I) or susceptible
(S). 
Infection events, i.e., $I+S \to I+I$, occur according to a Poisson process
with rate $\lambda \geq 0$.
Recovery events, i.e., $I \to S$, obey
a spontaneous Poisson process with rate $\mu$. 
We set $\mu = 1$,
with no loss of
generality. 
The configuration where all nodes are in the S state is an absorbing configuration: once such a configuration is reached the system will remain in it forever, as no new infections can arise. 
In a network of infinite size,
if the dynamics is 
started from a configuration different from the absorbing one, a critical value $\lambda_c$
of the control parameter $\lambda$ separates a phase where the system
unavoidably reaches the absorbing configuration
(i.e., $\lambda \leq \lambda_c$) from a phase where a stationary state with a finite
fraction of infected nodes exists ( $\lambda > \lambda_c$).

On a network with finite size $N$, the SIS model does not undergo a true phase transition. As long as the rate of infection $\lambda$ is finite, the system necessarily ends up in the absorbing configuration in a finite time. Nonetheless, it is still possible to identify a pseudo-critical point $\lambda_c(N)$, distinguishing a phase where the system reaches the absorbing configuration in a time that, on average, grows at most logarithmically with the system size (i.e., $\lambda < \lambda_c(N)$), and a phase where the absorbing configuration is reached in an average time that grows exponentially with the system size (i.e., $\lambda > \lambda_c(N)$).

In this paper, we are interested in characterizing SIS spreading on networks of finite size $N$ in their pseudo-critical regime, i.e., $\lambda = \lambda_c(N)$. We perform a large-scale numerical analysis and develop a theoretical approach valid for a specific initial setup where
$i_0 \ll N$ infected nodes are surrounded
by a completely susceptible system. 
Our main goal is understanding in detail the behavior of
the survival probability $S(t)$, i.e., the
probability that the system has not yet reached the absorbing configuration 
by time $t$, connected to the duration distribution as $P(t) = -dS(t)/dt$.
To avoid verbosity, we refer to the critical regime even if a network has finite size. Also, we make use of the compact notation $\lambda_c$ instead of $\lambda_c(N)$ to indicate the pseudo-critical point
of a network with finite size $N$.

\subsection*{Networks}

As interaction patterns, we consider networks built
according to the 
CM~\cite{molloy1995critical}.
The properties of a network generated according to the CM are specified by its degree distribution,
i.e., the probability $P(k)$ that a randomly selected node has degree $k$.
Each quenched network realization of the CM is obtained by sampling the
degree sequence 
$\vec{k} = (k_1, k_2, \ldots, k_N)$
from $P(k)$ and pairing nodes in a completely
random manner, while preserving the degree sequence.
The only constraint we set on the pairing procedure is that 
multiedges and self-loops are prohibited.
The $q$-th sample moment of the degree sequence is given by 
\begin{equation}
    \langle k^q \rangle = \frac{1}{N} \, \sum_{i=1}^N (k_i)^q \; .
    \label{eq:moment}
\end{equation}
The annealed version of the CM model implies that a new pairing is
generated at each time step, still keeping the degree sequence fixed.
In the annealed scenario one can usefully define the annealed adjacency matrix as the 
average over all possible pairings: $\bar{A}(k, k')=k k'/(\av{k}N)$. Its elements equal the
probability that a node of degree $k$
is connected to a node of degree $k'$~\cite{Dorogovtsev2008}.

In this work, we focus on power-law degree sequences generated by selecting random variates from the distribution 
\begin{equation}
P(k) \sim \left\{
\begin{array}{ll}
k^{-\gamma} & \textrm{ if } \, k \in 
[k_{\min}, k_{\max}]
\\
0 & \textrm{ otherwise }
\end{array}
\right.
\; .
    \label{eq:pl}
\end{equation}
In the following, without loss of generality, we set $k_{\min}= 3$, however,
results are qualitatively similar for $k_{\min} \geq 3$, as long as $k_{\min}$ is independent of $N$.
We assume that the maximum degree $k_{\max}$ is growing as a power of $N$, i.e., 
\begin{equation}
    k_{\max} = N^{1/\omega} \; ,
    \label{eq:kmax}
\end{equation}
with 
\begin{equation}
\omega \geq \max\{2, \gamma - 1\} \; .
    \label{eq:omega}
\end{equation}
This specific setting of the CM model is known as the uncorrelated configuration model (UCM)~\cite{Catanzaro2005}. Imposing the constraints of Eqs.~(\ref{eq:kmax}) and (\ref{eq:omega}) provides several advantages. For example,
one can safely assume that the effective maximum degree observed in the sequence sampled from $P(k)$ is actually $k_{\max}$ if $N$ is sufficiently large~\cite{boguna2009langevin}. Also, quenched UCM networks have negligible degree-degree correlations. 

Using the above constraints, one finds that the $q$-th moment of a large but finite network scales approximately as
\begin{equation}
\langle k^q \rangle \sim 
\left\{
\begin{array}{ll}
\textrm{const.} &  \textrm{ if } \, q < \gamma - 1
\\
\log k_{\max} =  1/\omega \, \log N & \textrm{ if } \, q = \gamma - 1
\\
k_{\max}^{q - \gamma} = N^{(q - \gamma) /\omega} &  \textrm{ if } \, q > \gamma - 1
\end{array}
\right. \; .
    \label{eq:mom_scaling}
\end{equation}
For $2 < \gamma \leq 3$, 
the power-law degree distribution of Eq.~(\ref{eq:pl}) has diverging second moment, i.e.,  $\langle k^2 \rangle \to \infty$ for $N \to \infty$.
We refer to networks in this class as heterogeneous networks. For $\gamma > 3$ instead, $\langle k^2 \rangle$ is finite, and we refer to this class as homogeneous networks.

\section{Summary of the results}
\label{sec:summary}

We consider SIS dynamics on a finite network of size $N$ initialized with $i_0 \ll N$ infected nodes. The sum of the degrees of the $i_0$ seeds is $k_0$. The survival function $S(t)$ turns out to be characterized by three main regimes describing respectively short, intermediate and long avalanches. We denote the point of transition between the first two regimes as $t^*$ and the point of transition between the second and third regime as $t_{\times}$. The transitions between the various regimes are not sharp, rather the survival function displays smooth crossover behaviors. 
This qualitative behavior is valid irrespectively of the  exponent $\gamma$ of the degree distribution in Eq.~(\ref{eq:pl}). Also, both the quenched and the annealed versions of the UCM exhibit this qualitative behavior. However, fundamental differences between homogeneous and heterogeneous networks exist as the actual values of the transition points depend on the degree exponent $\gamma$.

\subsection{Regime of short avalanches}
    
For $t < t^*$, in the regime of short avalanches, the survival function of each of the $i_0$ avalanches is characterized by an exponential decay.
This is due the immediate recovery of each of the $i_0$ initial spreaders.
The characteristic time scale of the exponential decay is equal to 1 because we set the rate of recovery as $\mu = 1$. 
If the $i_0$ avalanches are small enough not to interact one with the other, we can write $S(t) = 1 - (1-e^{-t})^{i_0}$, 
i.e., the survival probability equals
the probability that at least one of the $i_0$ avalanches is still active
at time $t$. 
Such expression behaves as
    \begin{equation}
        S(t) \sim i_0 \, e^{-t}
        \; .
        \label{eq:reg1}
    \end{equation}
if $t$ is sufficiently large. In particular, the above 
expression must be at most 1 so it can hold only if
$t \geq \log(i_0)$.

The relative weight of the regime of short avalanches compared to the other regimes valid for longer avalanches, which ultimately determines the transition point $t^*$, strongly depends on the initial condition of the dynamics and the topology of the underlying network. 
If $\gamma >3$, then $t^*$ does not display any dependence on the network size; if, instead, $2 < \gamma \leq 3$, a logarithmic divergence of $t^*$ with the system size appears. For $2 < \gamma \leq 3$, the 
fraction of trajectories that ends in
this regime also depends on the initial condition. If the spreading is initiated by spreaders with a sufficiently large $k_0$, then $t^* \simeq 0$ and the regime is barely visible, while $t^*$ grows
as $k_0$ diminishes.

\subsection{Regime of intermediate avalanches}

The range $t^* < t < t_{\times}$  is known as the adiabatic regime~\cite{boguna2009langevin, radicchi2020classes}. This regime describes avalanches that are sufficiently long to have lost any memory of the initial conditions. The survival function decays as
\begin{equation}
    S(t) \sim t^{-1} \; .
    \label{eq:reg2}
\end{equation}  
The power-law decay with exponent $-1$ is typical of critical spreading on networks with finite second moment of the degree distribution~\cite{di2017simple,radicchi2020classes,larremore2012statistical}. 
Here we find that the same decay describes intermediate avalanches in 
finite networks characterized by degree distributions with diverging second moment.

 \subsection{Regime of long avalanches}  
    
Finally, avalanches that last $t > t_{\times}$ are sufficiently large to feel the finite size of the network. An exponential cutoff characterizes the decay of the survival function in this regime, i.e.,

\begin{equation}
        S(t) \sim e^{- t / t_{\times}} \; .
    \end{equation} 
    
In particular, we have that the typical time of the exponential cutoff  diverges as a power of the network size, i.e.,
\begin{equation}
        t_{\times}(N) \sim N^{1/\sigma \nu} \; .
        \label{eq:reg3}
    \end{equation}
The specific value of  $\sigma \nu$ depends on $\gamma$ and $\omega$ only if $\gamma \leq 3$; $\sigma \nu = 2$ otherwise.

\section{Numerical results for annealed networks}
\label{sec:annealed_results}

In this section, we consider the case of uncorrelated annealed networks~\cite{boguna2009langevin}.
We consider the critical dynamics, setting $\lambda=\lambda_c=\av{k}/\av{k^2}$,
 the exact epidemic threshold for uncorrelated networks.
 We consider networks of finite size, and therefore the scaling of $\av{k}$ and $\av{k^2}$ with  $N$, is the one in Eq.~(\ref{eq:mom_scaling}).
It follows that, for $2 < \gamma \leq 3$, $\lambda_c \to 0$ for $N \to \infty$; instead, $\lambda_c$ does not vanish for $N \to \infty$ if $\gamma > 3$~\cite{PastorSatorras2015}.

\subsection{Homogeneous networks}

\begin{figure}
\begin{center}
\includegraphics[width=0.45\textwidth]{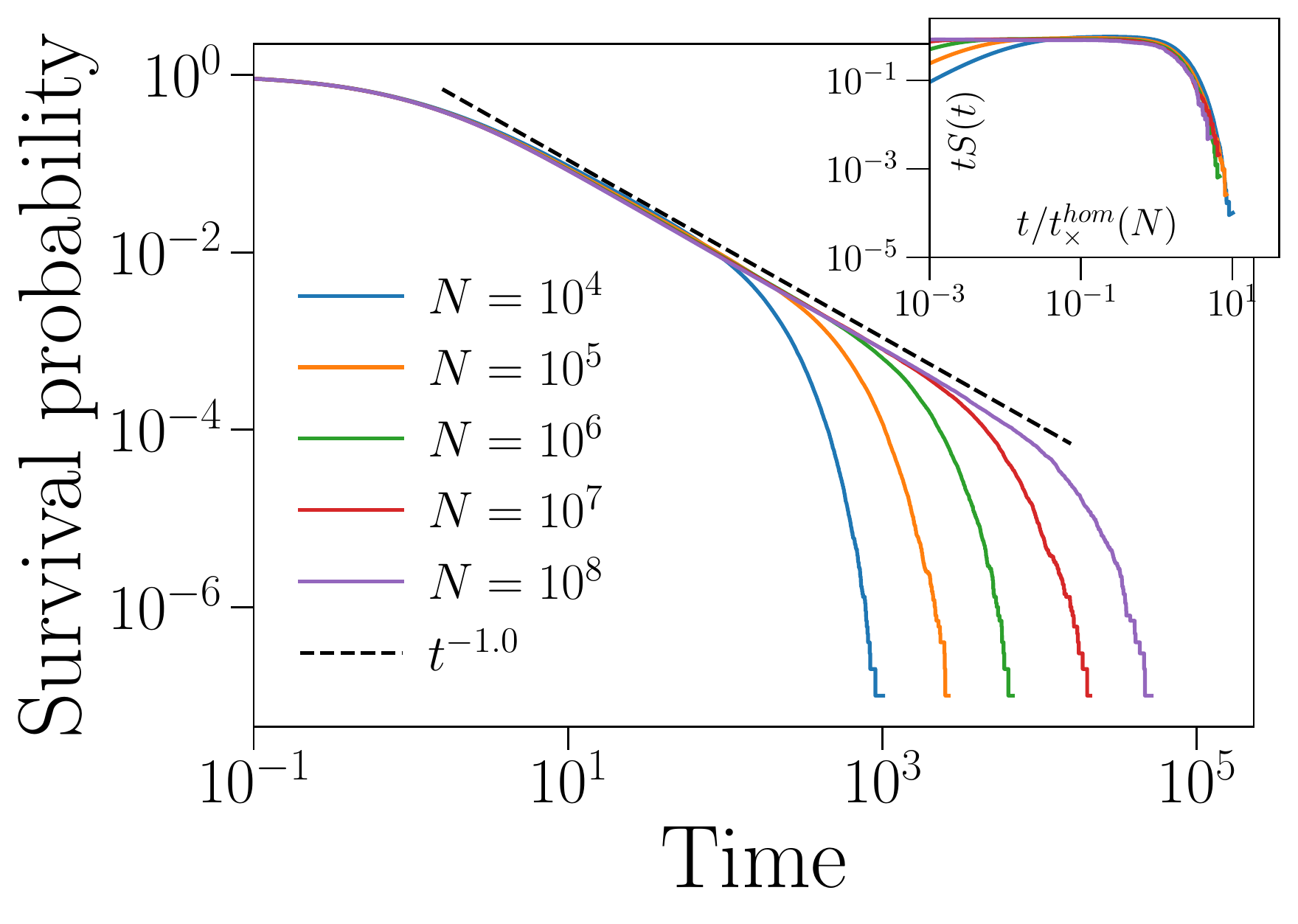} 
\end{center}
\caption{Survival probability for critical SIS avalanches in homogeneous annealed networks. We set $\gamma=5.3$,
  $\omega=4.3$, $k_0 = \floor{k_{min}(\gamma-1)/(\gamma-2)} =
  \floor{\av{k}}$ and we consider $10^7$ realizations of the process
  for each value of $N$. 
  Different curves show the survival probability $S(t)$ as $N$ is varied. 
  The black dashed line scales as
  $t^{-1}$.  In the inset, the same data as in the main panel are rescaled to make the various curves collapse one on the top of the other. The abscissa values are
  rescaled as $t/t_{\times}(N)$ with $t_{\times}(N) = N^{1/2}$, and the ordinate values are rescaled as $t S(t)$.
  }
\label{fig:homog}
\end{figure}

Figure~\ref{fig:homog} shows results obtained by simulating critical SIS dynamics
on homogeneous networks generated via the CM with
$P(k) \sim k^{-\gamma}$, $\gamma=5.3$ and $\omega= \gamma -1 = 4.3$.
All realizations have been initiated with a single
infected node, i.e., $i_0=1$, of degree
$k_0 = \floor{k_{min}(\gamma-1)/(\gamma-2)} = \floor{\av{k}}$, where $\floor{\cdot}$ is the floor function. This choice is motivated by the need of having an initial condition that is independent of the system size. 

The three regimes anticipated in the summary above are clearly visible. Given the homogeneity of the network, 
short avalanches do not display any dependence on the network size.  This is the standard behavior of the continuous-time BP~\cite{garciamillan2018field} and hence it is not surprising to observe it on homogeneous networks. In particular, we observe $t^* \simeq 1$. In the regime of intermediate avalanches, 
the survival probability shows the power-law
decay of Eq.~(\ref{eq:reg1}). 
The exponential cutoff emerges after a time
\begin{equation}
t_{\times}^{hom}(N) \sim N^{1/2} \; .
\label{eq:tc_homog}
\end{equation}
This is apparent from the finite-size scaling analysis shown in 
the inset of Fig.~\ref{fig:homog}.  This fact is consistent with what one finds  for other critical spreading processes on homogeneous networks via a standard mean-field description~\cite{henkel2008non}. 

The standard mean-field picture emerging from the analysis of the SIS model on homogeneous networks is 
further confirmed by the fact that the probability
distribution of the avalanche size
scales with exponent $3/2$ and shows a cutoff diverging linearly
with the system size, see 
Appendix A.
Both these results are consequences of the behaviour of $S(t)$
and of the scaling 
law that relates the average
avalanche size with its duration~\cite{di2017simple}.

\subsection{Heterogeneous networks}

\begin{figure*}
\begin{center}
\includegraphics[width=0.85\textwidth]{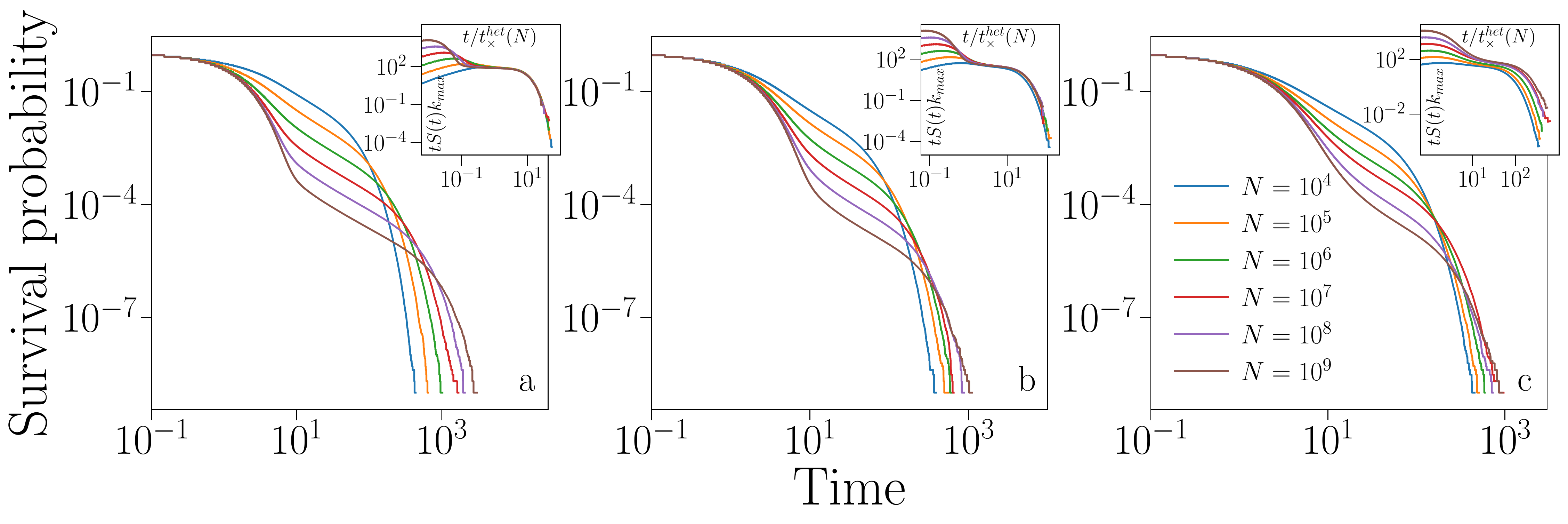}    
\end{center}
\caption{Survival probability for critical SIS avalanches in heterogeneous annealed networks. We set
$\omega=2$, $k_0 = \floor{k_{min}(\gamma-1)/(\gamma-2)} = \floor{\av{k}}$
and we consider $10^9$ realizations of the SIS process
for each value of the system size $N$. 
(a) We display the survival probability for $\gamma=2.1$. The inset displays the same data as in the main panel with the abscissa rescaled as 
$t/t_{\times}(N)$ and with the ordinate rescaled as $t S(t) k_{\max}$.
(b) Same as in panel (a), but for $\gamma=2.5$.
(c) Same as in panel (a), but for $\gamma=2.9$.
}
\label{fig:heterog}
\end{figure*}

Results for heterogeneous networks with degree exponent
$2 < \gamma \leq 3$
are shown in Figure~\ref{fig:heterog}. Also in this case, all realizations have been initiated with $i_0=1$ infected node of degree
$k_0 = \floor{k_{min}(\gamma-1)/(\gamma-2)} = \floor{\av{k}}$.
Despite Eq.~(\ref{eq:mom_scaling})
predicts $\av{k} \sim$ const. for any $\gamma>2$, in small systems such as the ones that can be considered numerically, $\av{k}$ has not yet reached its limit value and displays a slow dependence on $N$.
The three qualitative regimes are still present, but
the crossovers between the various regimes
strongly depend on $\omega$ and $\gamma$, in stark contrast with the homogeneous problem.

Short avalanches are described by the exponential decay of Eq.~(\ref{eq:reg1}). 
We observe that 
$t^*$ grows as $N$ increases. The reason for such a behavior is quite intuitive. 
The critical value of the epidemic threshold goes to zero as the system size increases. However, the initial condition of the dynamics is unaffected by the network size, as we are still considering spreading processes initiated by $i_0=1$ nodes with degree $k_0$. Many configurations generate short avalanches just because the probability of observing spreading events instead of recoveries becomes negligible as the size of the network increases. For example, the probability that the first event is a spreading event is $k_0 \lambda_c / (1 + k_0 \lambda_c) $, which clearly goes to zero as the system size increases. 

\begin{figure*}[!htb]
\begin{center}
\includegraphics[width=0.85\textwidth]{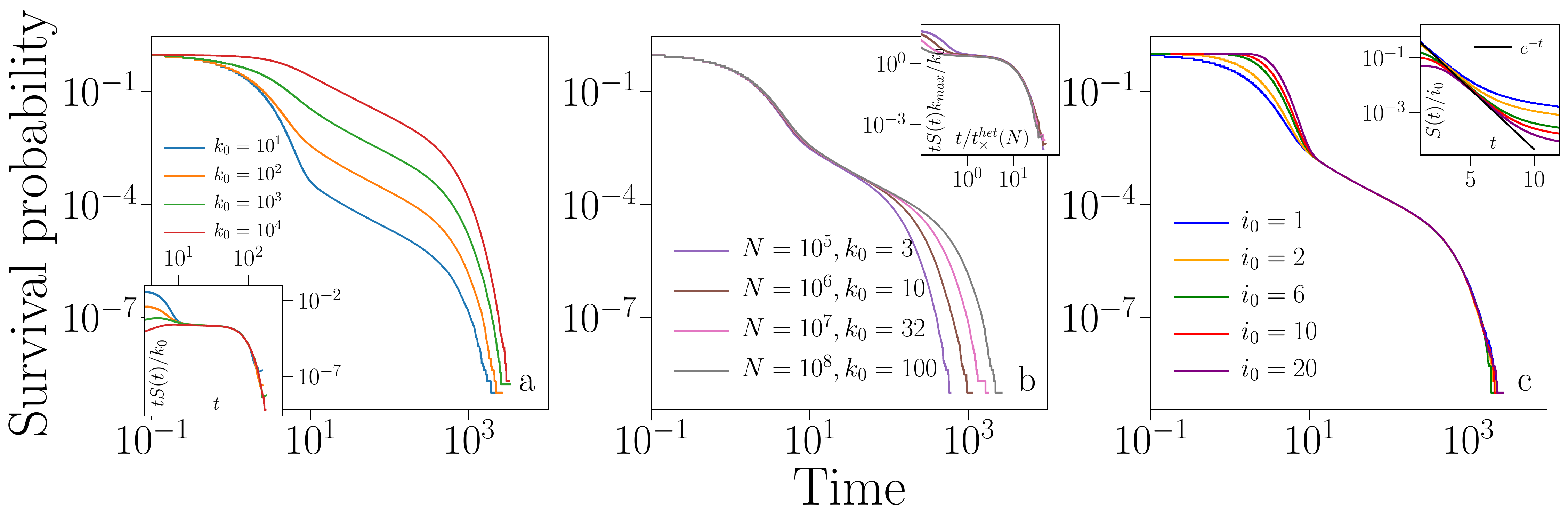}   
\end{center}
\caption{Survival probability for critical SIS avalanches in heterogeneous annealed networks. We set $\omega=2$, $\gamma=2.1$ and we consider $10^9$ 
realizations of the process for each curve. The process is started from $i_0$ seeds; the sum of the degrees of these seeds is $k_0$.
(a) The process is started from $i_0=1$. Different curves correspond to different values of $k_0$. Here the network size is $N=10^8$. 
The inset displays the same data as in the panel with the ordinate rescaled as 
$t S(t) / k_0$.
(b) We start the process from $i_0=1$ seed. We consider different $k_0$ values, but also different network sizes $N$. The values of these two parameters are chosen such that the ratio $k_{\max} / k_0 = 100$.
The inset displays the same data as in the main panel with the abscissa rescaled as 
$t/t_{\times}(N)$, with $t_{\times}(N)$ given in Eq.~(\ref{eq:tc_hetero}). We rescale also the ordinate as $t S(t) k_{\max} / k_0$.
(c) We start the process from variable values of $i_0$ all with the same
degree, but we keep $k_0 = 60$.
The inset displays the same data as in the main panel with the ordinate rescaled as $S(t)/i_0$. The solid black line is the exponential decay with unitary rate $e^{-t}$.}
\label{fig:InitCond}
\end{figure*}

This physically intuitive picture can be made more precise by studying the
dependence of the survival function in the regime of short avalanches 
on the initial condition
of the dynamics. Our main finding in this regard is that 
the crossover from the exponential to the power-law decay starts when the survival probability
has dropped by an amount that we infer to scale as 
$k_{\max}/k_0 \sim N^{1/\omega}/k_0$.
This finding is deduced from the combination of the results of Figs.~\ref{fig:InitCond}a, ~\ref{fig:InitCond}b
and ~\ref{fig:InitCond}c. 
In Fig.~\ref{fig:InitCond}a, we start the dynamics from 
a single initial seed, i.e., $i_0=1$ with variable degree $k_0$. Simulations are all performed on the same network, thus $k_{\max}$ is constant.  
The drop of $S(t)$ is proportional to $1/k_0$, as apparent from the inset of Fig.~\ref{fig:InitCond}a. In Fig.~\ref{fig:InitCond}b instead, we consider networks of different size to keep the ratio $k_{\max}/k_0$ constant. In the regime of short avalanches, the different curves collapse with no rescaling. Finally, in Fig.~\ref{fig:InitCond}c, we vary the number of initially infected nodes $i_0$, but keep constant the sum of their degrees:  $k_0 = 60$.
The value of the initial drop of $S(t)$ is once again constant, as the ratio $k_{\max}/k_0$ is not varied. The value of the time $t^*$ appears the same for all settings.
For $t \geq \log(i_0)$, the
functional form of $S(t)$ depends linearly on $i_0$, as described by Eq.~(\ref{eq:reg1}).

The scaling $k_{\max}/k_0$ of the initial drop of the survival function $S(t)$ together with the known exponential decay characterizing $S(t)$ in the regime of short avalanches, i.e., Eq.~(\ref{eq:reg1}), and the imposed $N$ dependence of the maximum degree, i.e., Eq.~(\ref{eq:kmax}),
allow us to write that
\begin{equation}
    t^*(N) \sim \log \frac{k_{max}}{k_0} \sim \frac{1}{\omega} \log N - \log k_0\; ,
    \label{eq:tstar_heterog}
\end{equation}
thus a logarithmic growth of the time distinguishing short from intermediate avalanches.

Intermediate avalanches are perfectly described by the power law of Eq.~(\ref{eq:reg2}).
Our finite-size scaling analysis indicates that 
\begin{equation}
    t_{\times}^{het}(N)  \sim N^{(\omega + 1 - \gamma) / 2 \omega} \, .
    \label{eq:tc_hetero}
\end{equation}
The scaling of Eq.~(\ref{eq:tc_hetero}) is clearly different from the one predicted for homogeneous networks, i.e., Eq.~(\ref{eq:tc_homog}). It can be, however, linked to the same physical interpretation behind Eq.~(\ref{eq:tc_homog}) using the known fact that in the SIS model on annealed networks, at criticality, the subset of nodes with largest degree plays a crucial role~\cite{PastorSatorras2015}. For $2<\gamma \leq 3$, this set is subextensive, and contains the $N_h \sim N^ {1+ (1-\gamma)/\omega}$ nodes with largest degree~\cite{pastor2016distinct}. The subgraph of the $N_h$ nodes with largest degree is also known to be homogeneous, in the sense that 
the smallest, largest and average degree of this subgraph all scale in the same way with the system size, as in homogeneous networks~\cite{pastor2016distinct}.
As a matter of fact, we can interpret the finding of Eq.~(\ref{eq:tc_hetero}) as $t_{\times}^{het}(N)  \sim t_{\times}^{hom}(N_h) \sim N_h^{1/2}$, thus as the cutoff time valid for a homogeneous graph, i.e., Eq.~(\ref{eq:tc_homog}), with $N_h$ nodes. 
Note that Eqs.~(\ref{eq:tstar_heterog}) and~(\ref{eq:tc_hetero})
explicitly depend on $\omega$ [see also Eq.~(\ref{eq:kmax})]. The validity of these expressions as $\omega$
is varied is assessed in Appendix~\ref{app:omega}.

\section{Langevin theory for annealed networks}
\label{sec:langevin_main}

The striking differences between the two cases of homogeneous and heterogeneous networks
call for an analytical investigation. A natural choice for this
task is to develop a theory based on a Langevin equation for the order parameter. 
The general theory of stochastic processes~\cite{gardiner1985handbook} provides
with an exact protocol to compute $S(t)$ given a Langevin equation for the order parameter
and such a protocol has been successfully employed to describe the behaviour of avalanches
in the CP~\cite{boguna2009langevin}. Inspired by Ref.~\cite{boguna2009langevin}, we develop a theoretical approach for the SIS model on annealed networks.

For annealed networks the degree sequence fully determines the network
properties: nodes with the same degree have exactly the same
dynamical behavior.
It is therefore convenient to define the variables $n_k(t)$ and $\rho_k(t) = n_k(t)/ [NP(k)]$
which represent, respectively, the number of infected nodes
with degree $k$ at time $t$ and the probability that a randomly
selected node among those with degree $k$ is infected at time $t$.
The epidemic phase transition can be described through the order parameter
$\rho = \sum_k n_k / N = \sum_k \rho_k P(k)$, representing the
probability that a random node is infected, or through the order
parameter $\Theta = \sum_k k \rho_k P(k) /\av{k}$, representing the
probability that a random neighbor of a node is infected.

The theoretical approach is based on the observation that the variables
$n_k$, and hence $\rho_k$, are sums of a large number of binary random variables,
which for annealed networks are independent~\cite{boguna2009langevin}.
Appealing to the central limit theorem, the temporal evolution
of these variables can be described by means of suitable Langevin equations.

Following closely the approach of Ref.~\cite{boguna2009langevin} valid for the CP, we find the Langevin 
equation valid for the critical SIS model is
\begin{equation}
\dot{\Theta} = - \frac{\av{k}\av{k^3}}{\av{k^2}^2} \Theta^2 +
 \sqrt{\frac{2 \Theta}{N} \frac{\av{k^3}}{\av{k}\av{k^2}} } \, \xi \, ,
\label{eq:Theta_Lang}
\end{equation}
where $\xi$ is a Gaussian white noise of zero mean and unit variance.
The derivation of this result is in appendix~\ref{sec:langevin}, and is based on three main hypotheses. We assume that the dynamics takes place on an annealed uncorrelated network, the variables $\rho_k$ evolve in time 
subjected to Gaussian white noise and, importantly, the adiabatic approximation is valid.

Such an approximation consists in replacing the value of the microscopic variables $\rho_k$ in the equation for the evolution of the order parameter $\Theta$ with quasi-stationary, $\Theta$-dependent, values, obtained by setting the time derivative to zero in their evolution equations. This is based on the fact that $\Theta$ evolves slowly at criticality, while the $\rho_k$ relax exponentially fast. 
(see Appendix~\ref{sec:langevin}). Only  
sufficiently long avalanches
may be correctly described by our theory. In particular, the Langevin theory 
can not explain 
the regime of short avalanches, where the exponential decay is observed. 
This happens because the adiabatic approximation is not yet valid. 


Eq.~(\ref{eq:Theta_Lang}) can be recast as a partial differential equation for $S(t)$, which  
in turn can be used to compute the finite-size 
properties of the survival function. Explicit calculations are shown in Appendix~\ref{app:surv}. 
At criticality, the theory predicts 
\begin{equation}
    S(t) = 1 - e^{-\bar{t} / t}
    \label{eq:St}
\end{equation}
where 
\begin{equation}
    \bar{t} = k_0 \av{k^2} / \av{k^3} 
    \label{eq:bar_t} \; ,
\end{equation} 
and with a cutoff for long times, due to the finite
size of the network, scaling as
\begin{equation}
    t_{\times}(N) \sim \sqrt{N \av{k^2}} \left(\frac{\av{k^2}}{\av{k^3}} \right)^2 \, .
    \label{eq:tc}
\end{equation}

For $t \gg \bar{t}$, Eq.~(\ref{eq:St}) predicts that $S(t) \sim t^{-1}$. This prediction is valid for all networks, homogeneous and heterogeneous, and is in  agreement with the results of numerical simulations. We stress that $\bar{t}$ does not represent the same quantity as $t^*$: the regime of short avalanches is not described under the adiabatic approximation which we do not expect to be valid during the regime of short avalanches.
Indeed, the cumulative probability of Eq.~(\ref{eq:St}) equals one for $t=0$, as  expected for a conditional probability such as the present one. $S(t)$ in Eq.~(\ref{eq:St}) is conditioned
on $t$ being larger than the adiabatic relaxation time. The fact that $\bar{t}$ tends to zero
in heterogeneous networks signals that the power-law decay begins immediately after the adiabatic relaxation time while it happens after a time that grows with $k_0$ in homogeneous networks.

For homogeneous networks, 
Eq.~(\ref{eq:bar_t}) predicts
$\bar{t}$ of order $k_0$ and 
Eq.~(\ref{eq:tc}) tells us
that $t_{\times}(N) \sim N^{1/2}$. 
These predictions are perfectly consistent with the numerical results presented in 
section~\ref{sec:summary}.

Predictions, however, are only partially in agreement with numerical simulations for heterogeneous networks. 
The scaling $\bar{t} \sim k_0 / k_{\max}$ predicted in Eq.~(\ref{eq:bar_t}) is consistent with the fact that $t S(t) k_{\max} / k_0$ is constant, see Fig.~\ref{fig:heterog} and 
Fig.~\ref{fig:InitCond}b. 
The theory, however, incorrectly predicts 
the scaling of the cutoff for large times $t_{\times}$. Indeed, for $2 < \gamma \leq 3$ both
$\av{k^2}$ and $\av{k^3}$ diverge as described in Eq.~(\ref{eq:mom_scaling}), 
and Eq.~(\ref{eq:tc}) gives
$t_{\times}(N) \sim N^{(\omega-\gamma-1)/2 \omega}$,
which diverges with $N$ only if $\omega > \gamma+1$ and 
limits to zero otherwise. 
This latter finding is not only in stark contrast with the results of the numerical simulations, but is also unphysical, denoting some underlying shortcoming of the theory. We believe that the
Langevin approach fails because it assumes the 
whole system to be involved in the critical dynamics and it does not capture the special role played by the subextensive graph of nodes with highest degree, as explained in the previous section. Hence, we can
recover a correct scaling of $t_{\times}$ by heuristically replacing $N$ with $N_h$ and using the scaling of the homogeneous problem. 


\section{Numerical results for quenched networks}
\label{sec:quench_results}

\begin{figure*}[!htb]
\begin{center}
\includegraphics[width=0.85\textwidth]{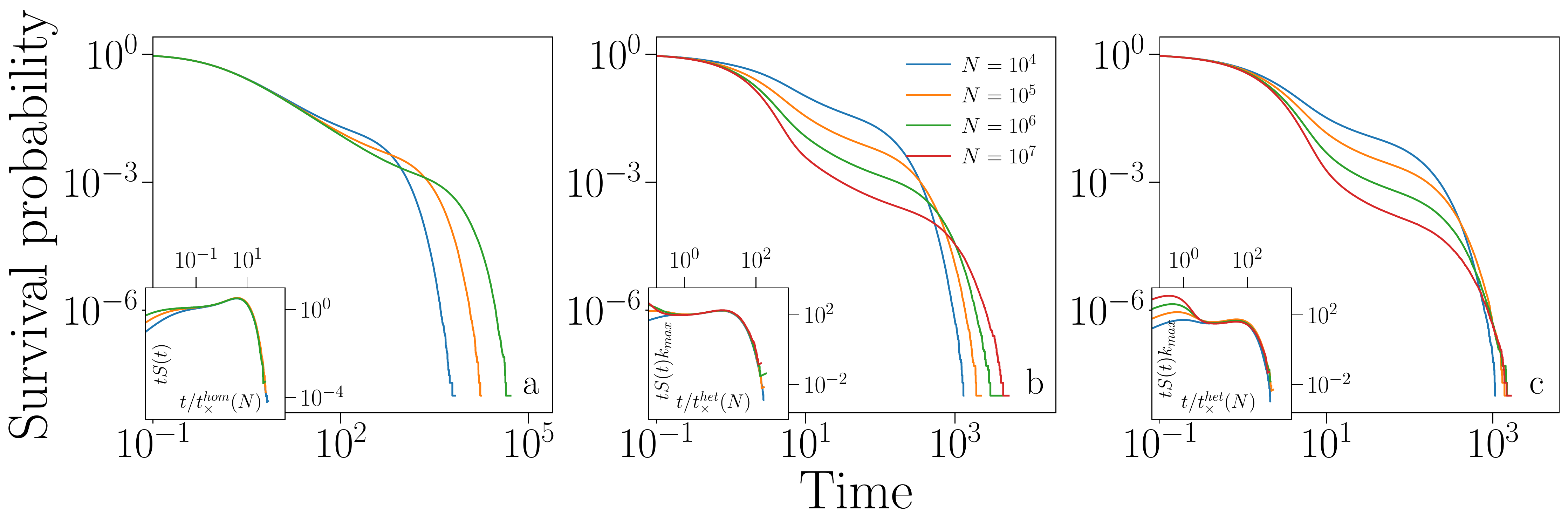}    
\end{center}
\caption{Survival probability for critical SIS avalanches in quenched networks. 
We set $k_0 = \floor{k_{min}(\gamma-1)/(\gamma-2)} = \floor{\av{k}}$.
(a) Simulations are performed on a random regular graph with degree $k=10$. 
We consider $10^8$ realizations of the spreading process
for each value of $N$. The inset displays the same data as the main panel with the abscissa rescaled as  $t/\bar{t}(N)$ and with the ordinate rescaled as $t S(t)$.
(b) Same as in panel (a) but on configuration model networks with $\gamma=2.1$ and 
$\omega=2$. We consider $10^9$ realizations of the SIS process for each value of $N$. The inset displays the same data as in the main panel with the abscissa rescaled as  $t/\bar{t}(N_h)$ and with the ordinate rescaled as 
$t S(t) k_{\max}$. (c) Same as in (b), but for $\gamma=2.7$ and 
$\omega=2$.
}
\label{fig:quench}
\end{figure*}

The comprehension of the phenomenology on annealed networks allows us to
interpret what happens in spreading experiments on quenched networks.

The main limitation to study large-scale networks is the 
computational cost of estimating the critical threshold of a network. At odds with the case of annealed networks, $\lambda_c$  is not simply given by the ratio between first and second moment of the degree sequence, rather it 
requires
to be numerically estimated~\cite{Ferreira2012}.
We use the standard practice of identifying $\lambda_c$ as the $\lambda$ value corresponding to maximum of the susceptibility associated to $\rho$, see 
Appendix~\ref{app:suscept}. Numerical estimates of $\lambda_c$ require high accuracy, up to six decimal digits\footnote{The number of digits required to have satisfactory results depends on the network.}, and these estimations are computationally demanding even though the maximization is performed using the efficient quasi-stationary method~\cite{Ferreira2012}. We have been able to study networks of size at most $N=10^7$.
Once, the value of $\lambda_c$ is estimated, 
spreading experiments for $\lambda=\lambda_c$ can be performed at a relatively
small computational cost. 

Results of our analysis are shown in Fig.~\ref{fig:quench}.
We first validate our method 
on random regular graphs, 
whose phenomenology is expected to be
perfectly consistent with the standard mean-field picture, i.e., consistent with the
predictions of the Langevin theory for homogeneous networks and the findings for homogeneous annealed networks (Fig.~\ref{fig:homog}).
Numerical results, shown in Figure~\ref{fig:quench}a, confirm this expectation
as $S(t)$ obeys Eq.~(\ref{eq:St}) and the cutoff
is given by Eq.~(\ref{eq:tc}).
Interestingly, the scaling function displays a hump just before the cutoff.
The hump is sufficiently
large to be observed even in the non-rescaled data and hinders the
direct observation of the exponent $-1$ for relatively small systems,
a fact that often happens when scaling functions display
large peaks~\cite{sethna2001crackling}.

The results for small values of $\gamma$ are perfectly analogous to those obtained for heterogeneous annealed networks. For example, in the case $\gamma=2.1$, Fig.~\ref{fig:quench}b, one finds that the
rescaling used for annealed networks leads to the clean data collapse
shown in the inset. This finding is not surprising, as the theory of annealed networks suits well SIS on quenched networks as long as $\gamma < 5/2$~\cite{Ferreira2012,pastor2016distinct}.
In Appendix~\ref{app:suscept} we show that the maximization of the susceptibility is a necessary step in order to obtain clean-cut results as the ones shown in Fig.~\ref{fig:quench}.

The rescaling used for annealed networks works quite well 
also for $\gamma=2.7$ (see Fig.~\ref{fig:quench}c). 
Also in this case we find
that the drop of $S(t)$ at small $t$ is proportional to $k_{\max}/k_0$ and
the cutoff is again given by Eq.~(\ref{eq:tc_hetero}). However, this 
agreement is valid for the relatively small network sizes we can simulate, i.e.,  $N \le 10^7$.
Based on the findings of Refs.~\cite{Ferreira2012,pastor2016distinct, Castellano2020}, we expect that, for any value $\gamma>5/2$, the theory for annealed networks is no longer able to describe critical SIS avalanches on sufficiently large quenched networks, and that a scaling different from Eq.~(\ref{eq:tc_hetero}) holds for $N \to \infty$.

\section{Conclusions}
\label{sec:conclusions}

In this paper we have performed a thorough analysis of SIS critical avalanche dynamics
on random networks
of finite size.
While for annealed homogeneous networks the theoretical understanding
is complete and quantitative, when the dynamics takes place on annealed heterogeneous
substrates we have to resort to heuristic arguments to correct some shortcomings of the Langevin
approach. In particular, it is necessary to assume that nodes play different roles
depending on whether they are part of the subextensive subgraph of size $N_h$ which
includes high-degree nodes. Such a network subset is where 
the principal eigenvector of the adjacency matrix gets localized~\cite{pastor2016distinct}.
For stationary properties of the SIS dynamics, this localization determines the position
of the epidemic threshold, but it does not affect the probability that a node is infected,
which is proportional to the degree for any value of $k$.
Our results show instead that for nonstationary properties, such as avalanche statistics,
localization matters: nodes outside the subset play no role in the evolution of long
avalanches and as a consequence the Langevin approach, which treats all nodes on the
same footing, fails.

This failure is in stark contrast with what happens for the CP.
Physically, the CP differs from the SIS model as node $i$ spreads activity 
with rate $\lambda$ regardless of its degree, rather than spreading with rate
$\lambda k_i$.
As a consequence, the critical point of the CP is $\lambda_c^{CP}=1$, 
i.e., is network independent while the critical point
of the SIS is not and, in particular, it vanishes in heterogeneous networks. 
However, the structure of the Langevin equations describing the two processes 
is quite similar. The equations for the variables
$\rho_k$ are the same, with the only difference that the order parameter $\Theta$ 
appearing in Eq.(~\ref{eq:rho_k_lang}) is replaced by the factor $\rho/\av{k}$ and 
Eq.(~\ref{eq:theta_init}) has the same structure as of the equation for $\rho$ in the CP, 
with the fundamental difference that the non-linear term involves
$\sum_k k P(k) \rho_k$ in the CP, while it involves $\sum_k k^2 P(k) \rho_k$ in the SIS.
This difference reflects the physically different spreading mechanisms of the two models.
Furthermore, the analogy between the 
equations for the $\rho_k$ and for the order parameters allows to perform the adiabatic approximation
in both cases and the stationary values of the $\rho_k$ have indeed the same form.
It follows that the closed-form expressions of the Langevin equations are the same for both processes.
The physical difference between the
two processes, however, is reflected by a different scaling of the drift and diffusion coefficients with
the system size. In particular, in the limit of small order parameter, the SIS model involves 
moments of $P(k)$ up to the third (see Eq.~(\ref{eq:Theta_Lang})) while the highest moment in the 
CP is the second. 
Once the Langevin equations are obtained, writing the partial differential equation for $S(t)$,
its limit for large $N$ and its integration for the calculation of $t_{\times}$ proceeds analogously 
for both models. It was found that the CP is always characterized by a 
$t^*$ that does not scale with the system size regardless
of the network (as $\lambda_c^{CP}$ is finite on any network). The cutoff
time $t_{\times}$, however, is correctly predicted for the CP but not for the SIS.
A solid mathematical framework to explore this regime is still to be found.

We have verified that, for $\gamma<5/2$, the avalanche statistics on quenched networks
is practically identical to the case of annealed networks, in agreement with what
occurs for stationary properties~\cite{Ferreira2012}.
For larger values of $\gamma$ the stationary properties of SIS on quenched networks
(such as the value of the epidemic threshold) 
are governed by the interplay of hubs which effectively
``interact at distance'' giving rise to long-range percolation patterns~\cite{Castellano2020}.
The investigation of how these highly nontrivial effects influence critical avalanche statistics
for quenched networks with $\gamma>5/2$ remains a very challenging open problem both from
the numerical and from the analytical point of view.

\begin{acknowledgements}
F.R. acknowledges support by the Army Research Office (W911NF-21-1-0194) and by the Air Force Office of Scientific Research (FA9550-21-1-0446). The funders had no role in study design, data collection and analysis, decision to publish, or any opinions, findings, and conclusions or recommendations expressed in the manuscript.
\end{acknowledgements}

\appendix
\section{Avalanche size distribution}

\begin{figure}
\begin{center}
\includegraphics[width=0.45\textwidth]{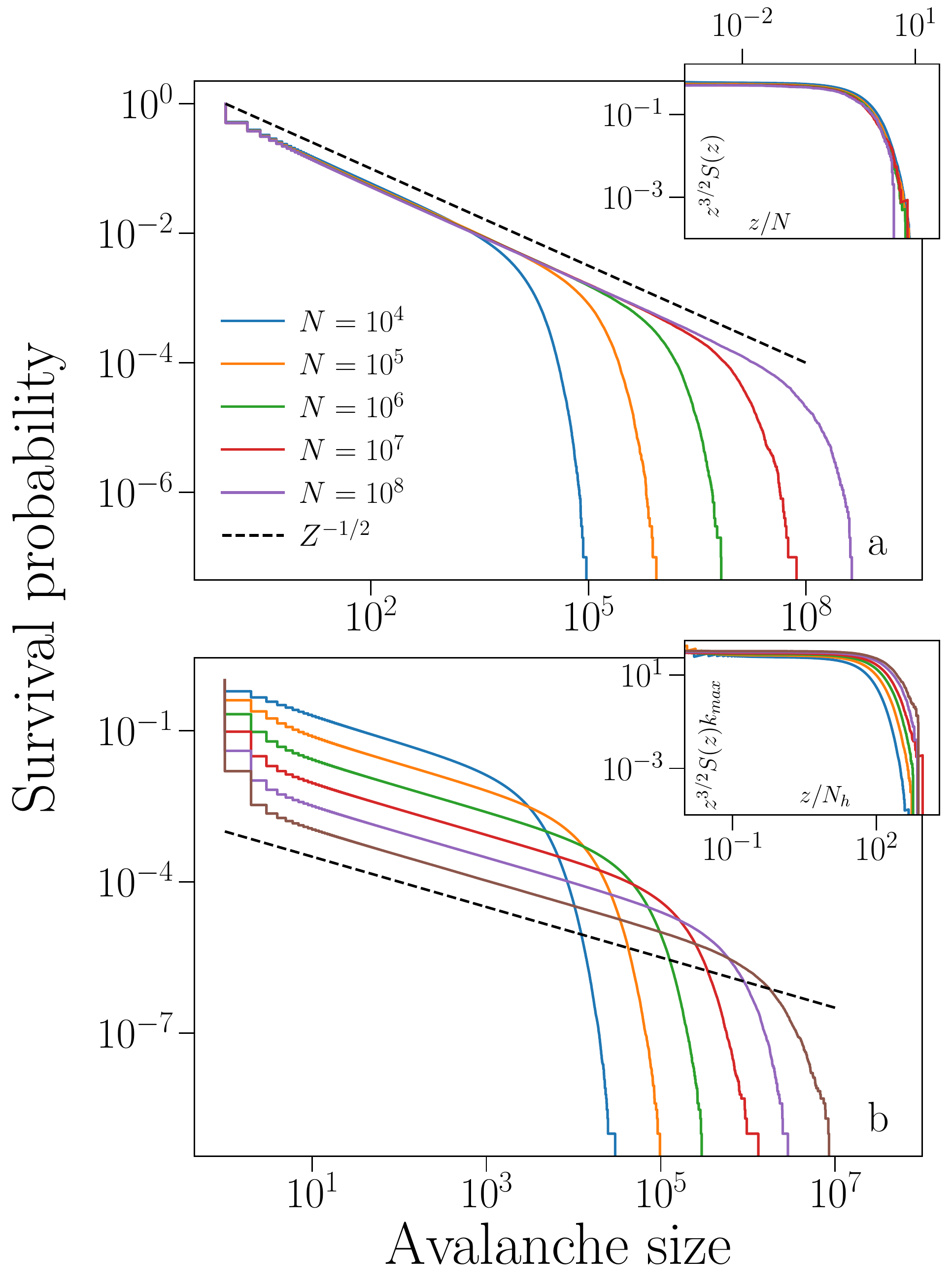} 
\end{center}
\caption{Survival probability of the avalanche size for critical SIS avalanches in annealed networks.
(a) We set $\gamma=5.3$, $\omega=4.3$, 
$k_0 = \floor{k_{min}(\gamma-1)/(\gamma-2)} = \floor{\av{k}}$
and we consider $10^7$ realizations of the SIS process
for each value of the system size $N$. 
The inset displays the same data as in the main panel with the abscissa rescaled as 
$z/N$ and with the ordinate rescaled as $z^{3/2} S(z)$.
(b) We set $\gamma=2.1$, $\omega=2$, 
$k_0 = \floor{k_{min}(\gamma-1)/(\gamma-2)} = \floor{\av{k}}$
and we consider $10^9$ realizations of the SIS process
for each value of the system size $N$. 
The inset displays the same data as in the main panel with the abscissa rescaled as 
$z/N_h$ and with the ordinate rescaled as $z^{3/2} P(z) k_c$.
}
\label{fig:size}
\end{figure}

Figure~\ref{fig:size} displays the avalanche size distribution obtained on
annealed networks. As explained in the main text, the power-law exponent
$\tau=3/2$ is easily understood from Eq.~(\ref{eq:reg2}) and from the
quadratic growth of the average avalanche size with the avalanche 
duration~\cite{di2017simple}.
The avalanche size cutoffs $z_{\times}$ for both classes of networks are understood again using $\theta = 2$ 
and in particular they are the square of the
cutoffs of the survival probability. The drop of $P(z)$ for small values of
$z$ in heterogeneous networks mirrors the exponential decay of the first temporal 
regime. During such a decay, avalanche size and duration scale 
linearly as, for short avalanches, an infection event increasing $z$ by 1 generally increases $t$ only of the healing time of the newly infected node, i.e., $1/\mu=1$ on average. Hence, for short avalanches $\av{z} \propto t / \mu$. It follows that
the rescaling is the same for both observables.

\section{The role of the upper cutoff of the degree distribution}
\label{app:omega}

\begin{figure}
\begin{center}
\includegraphics[width=0.45\textwidth]{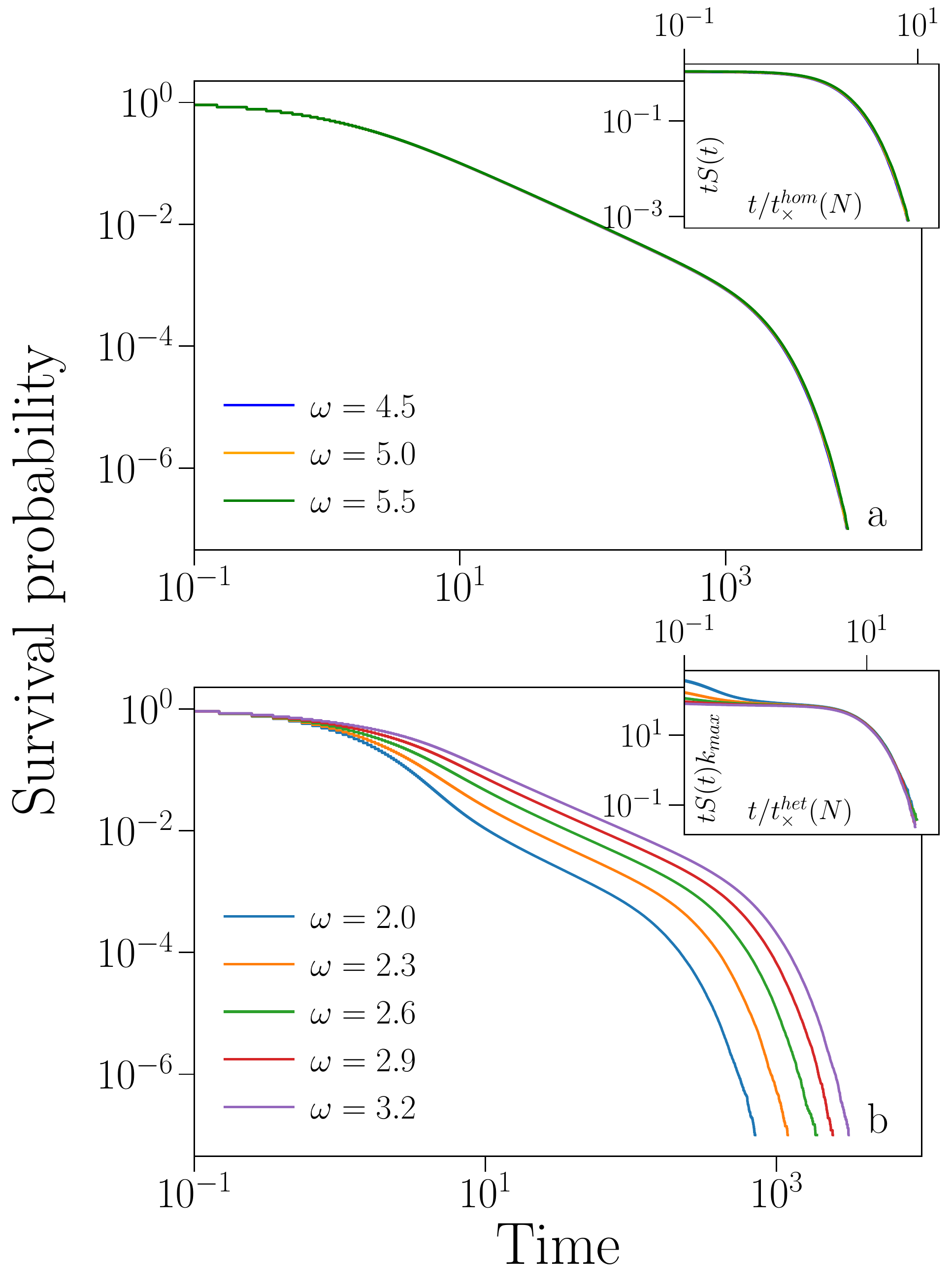} 
\end{center}
\caption{Survival probability for critical SIS avalanches in annealed networks.
(a) We set $\gamma=5.3$, $N=10^6$, 
$k_0 = \floor{k_{min}(\gamma-1)/(\gamma-2)} = \floor{\av{k}}$
and we consider $10^7$ realizations of the SIS process
for each value of $\omega$. 
The inset displays the same data as in the main panel with the abscissa rescaled as 
$t/t_c(N)$ and with the ordinate rescaled as $t S(t)$.
(b) We set $\gamma=2.1$, $N=10^6$, 
$k_0 = \floor{k_{min}(\gamma-1)/(\gamma-2)} = \floor{\av{k}}$
and we consider $10^7$ realizations of the SIS process
for each value of $\omega$. 
The inset displays the same data as in the main panel with the abscissa rescaled as 
$t/t_c(N_h)$ and with the ordinate rescaled as $t S(t) k_{max}$.
}
\label{fig:omega}
\end{figure}

The cutoff $t_{\times}$ is predicted to depend only on $N$ in
homogeneous networks, Eq.~(\ref{eq:tc_homog}), while it
explicitly
depends on $\gamma$ and $\omega$ in heterogeneous networks, Eq.~(\ref{eq:tc_hetero}). 
The relation between $t_{\times}$ and
$\gamma$ is explicitly assessed in Fig.~\ref{fig:heterog}.
Figure~\ref{fig:omega} verifies the relationship between $t_{\times}$
and $\omega$ by keeping all parameters fixed except for
$\omega$. For large $\gamma$ values, networks are 
homogeneous regardless of the $\omega$ value. 
For heterogeneous networks, in turn,
$N_h$ depends on $\gamma$, $\omega$ and $N$~\cite{dorogovtsev2006k}.

\section{Derivation of the Langevin equations}
\label{sec:langevin}

\begin{figure}
\begin{center}
\includegraphics[width=0.45\textwidth]{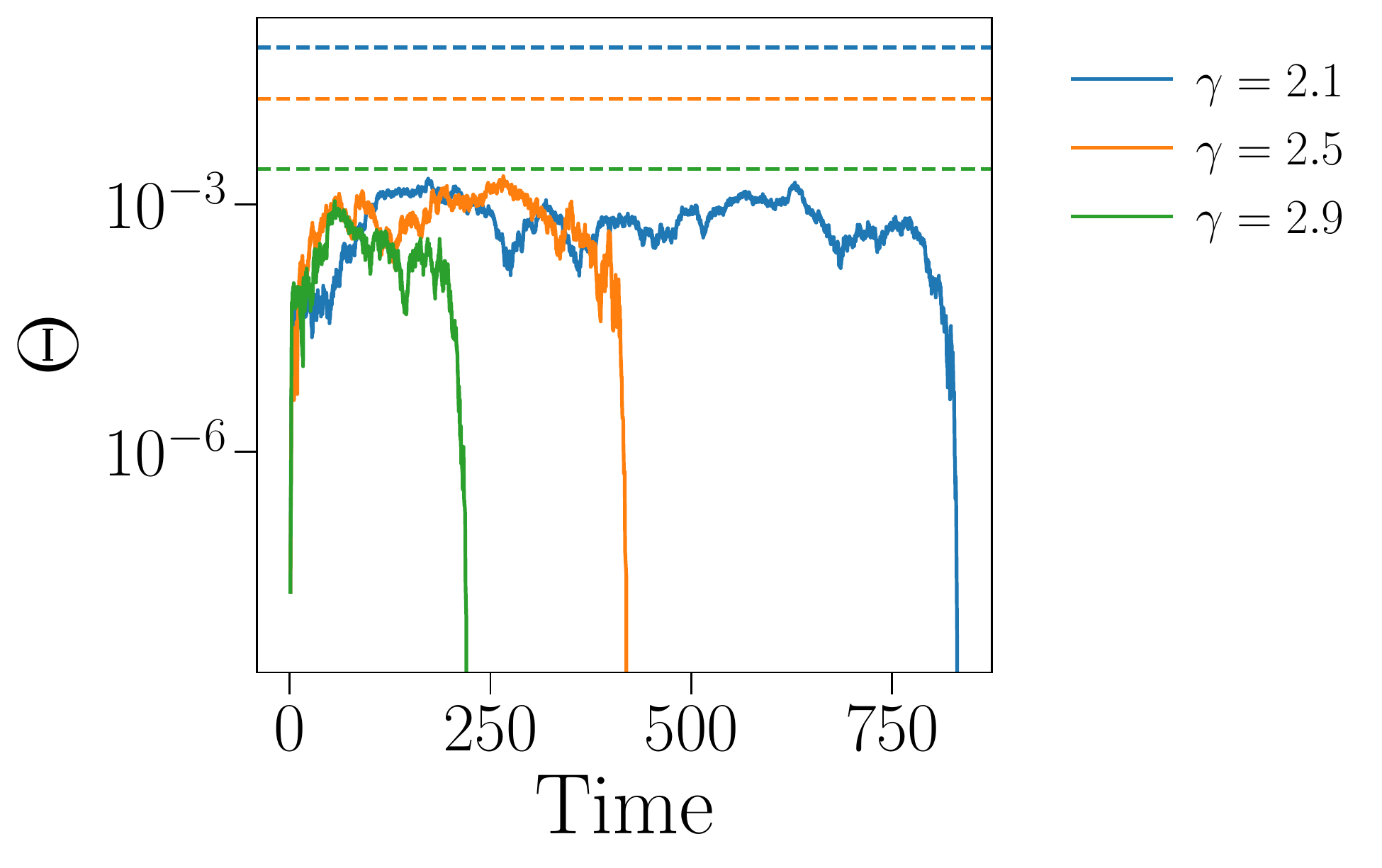} 
\end{center}
\caption{Trajectories of the order parameter $\Theta$. We use here
$N=10^8$ and $\omega=2$. Each realization is initialized with $k_0 = \floor{k_{min}(\gamma-1)/(\gamma-2)} = \floor{\av{k}}$. Solid lines correspond to
a single trajectory of the order parameter for different values of $\gamma$. 
Dashed lines correspond to the bound $1/\lambda_c k_{max}$. }
\label{fig:theta}
\end{figure}

The derivation of the Langevin equations proceeds in the same way as 
in~\cite{boguna2009langevin, radicchi2020classes}. The Langevin equations for the variables $n_k$ are
\begin{equation}
\begin{split}
	\dot{n}_k & = - n_k + \lambda \frac{k}{\av{k}} (1-\rho_k) \sum_{k'} k' P(k') n_{k'} \\
	& + \sqrt{ n_k + \lambda (1-\rho_k)  \frac{k}{\av{k}} \sum_{k'} k' P(k') n_{k'} }  \, \xi_k \, ,
\end{split}
\end{equation}
Diving by $N P(k)$ we obtain
\begin{equation}
\begin{split}
	\dot{\rho}_k & = - \rho_k + \lambda k (1-\rho_k) \Theta \\
	& + \sqrt{ \frac{1}{NP(k)} \Big[ \rho_k + \lambda k (1-\rho_k) \Theta \Big] }  \, \xi_k \, ,
	\label{eq:rho_k_lang}
\end{split}
\end{equation}
and using the definition of $\Theta$ 
\begin{equation}
\begin{split}
\dot{\Theta} & = \Theta \left(-1 + \lambda \sum_k \frac{k^2}{\av{k}} (1-\rho_k) P(k) \right) \\
& + \frac{1}{\av{k}}\sum_k k P(k) 
\sqrt{\frac{1}{N P(k)} \Big[ \rho_k +\lambda k (1-\rho_k) \Theta \Big]} \, \xi_k \, .
\label{eq:theta_init}
\end{split}
\end{equation}
Note that the leading order in Eq.~(\ref{eq:rho_k_lang}) is linear in $\rho_k$ regardless
of the value of $\lambda$, implying a fast exponential decay toward their asymptotic (average) values,
while linear terms are suppressed in Eq.~(\ref{eq:theta_init}) by setting $\lambda=\lambda_c$  and
hence $\Theta$ is a slowly varying variable at criticality. 
Therefore, the adiabatic approximation can be performed at the critical point.
Setting $\dot{\rho}_k=0$ and $\lambda=\lambda_c$ yield
\begin{equation}
\begin{split}
\rho_k & = \frac{\lambda_c k \Theta}{1 + \lambda_c k \Theta}  \\
& + \frac{1}{1+\lambda_c k \Theta} \sqrt{\frac{1}{NP(k)} \Big[ \rho_k +\lambda_c  k 
(1-\rho_k)  \Theta \Big]} \, \xi_k \, .
\label{c4}
\end{split}
\end{equation}
In the above expression, the stochastic term vanishes in the thermodynamic limit.
Therefore we can rewrite Eq.~\eqref{c4} by replacing $\rho_k$ in the square root with its leading (deterministic) term, so to express $\rho_k$ entirely as a function of $\Theta$, obtaining
\begin{equation}
\rho_k = \frac{\lambda_c k \Theta}{1 + \lambda_c k \Theta} + 
\sqrt{\frac{2\lambda_c k \Theta}{N P(k) (1+\lambda_c k \Theta)} } \; \xi_k .
\end{equation}
Replacing this expression in the deterministic term of  Eq.~(\ref{eq:theta_init}) gives
\begin{equation}
\begin{split}
& \Theta \left[-1 + \lambda_c \sum_k \frac{k^2 P(k)}{\av{k}}
\frac{1}{1+\lambda_c k \Theta} \right] - \\
& \lambda_c \Theta
\sum_k \frac{k^2P(k)}{\av{k}} \sqrt{\frac{1}{N P(k)} \frac{2\lambda_c k \Theta}
{1 + \lambda_c k \Theta}} \xi_k \; ,
\end{split}
\end{equation}
while the stochastic term
takes the form
\begin{equation}
\sum_k \frac{kP(k)}{\av{k}} \sqrt{\frac{1}{NP(k)} 
\frac{2 \lambda_c k \Theta}{1 + \lambda_c k \Theta}} \; \xi_k \, .
\end{equation}
The overall contribution to the new stochastic term in the equation for $\dot{\Theta}$ is
\begin{equation}
\begin{split}
& \sum_k \frac{kP(k)}{\av{k}} \sqrt{\frac{1}{NP(k)} 
\frac{2 \lambda_c k \Theta}{1 + \lambda_c k \Theta}} 
\left( 1 - \lambda_c k \Theta \right) \; \xi_k = \\ &
\sqrt{\sum_k \frac{k^2P(k)}{N\av{k}^2} \frac{2\lambda_c k \Theta}
{1+\lambda_c k \Theta}\Big( 1 - \lambda_c k \Theta \Big)^2} \; \xi \, .
\end{split}
\end{equation}
Note that, in inserting the sum into the square root, we also replace the
individual noises $\xi_k$ with an overall noise $\xi$ by means of the central
limit theorem. 
Defining $\Delta=\lambda_c \frac{\av{k^2}}{\av{k}} - 1$ and 
summing and subtracting
the quantity $\lambda_c \frac{\av{k^2}}{\av{k}} \Theta$, the closed-form 
of the Langevin equation for $\Theta$ is
\begin{equation}
	\dot{\Theta} = \Theta (\Delta - \lambda_c \Omega[\Theta]) + 
	\sqrt{\frac{2\lambda_c \Theta}{N} \Lambda[\Theta]} \; \xi \, ,
	\label{eq:theta_closed}
\end{equation}
where 
\begin{equation}
	\begin{dcases}
		\Omega[\Theta] = \sum_k \frac{k^2 P(k)}{\av{k}}
		\frac{\lambda_c k \Theta}{1+\lambda_c k \Theta} \\
		\Lambda[\Theta] = \sum_k \frac{k^3P(k)}{\av{k}^2}
		\frac{(1 - \lambda_c k \Theta)^2}{1+\lambda_c k \Theta} 
	\end{dcases} \, .
    \label{eq:OmegaLam}
\end{equation}
The two summations above can be easily performed if $\Theta \ll 1 / \lambda_c k_{max}$. 
Under this assumption we have $\Omega = \lambda_c \Theta \av{k^3} / \av{k}$ and 
$\Lambda = \av{k^3} / \av{k}^2$. 
Replacing these expressions in Eq.~(\ref{eq:theta_closed}) we obtain 
Eq.~(\ref{eq:Theta_Lang}). Figure~\ref{fig:theta} shows that
the assumption $\Theta \ll 1 / \lambda_c k_{max}$ holds if the dynamics is initialized with
sufficiently small values of $\Theta$, as we always do in our simulations.

\section{The survival probability function}
\label{app:surv}

From Eq.~(\ref{eq:Theta_Lang}) an equation for $S(t|\Theta_0)$
can be derived using standard techniques~\cite{gardiner1985handbook}
\begin{equation}
    \begin{split}
    \frac{\partial}{\partial t} S(t|\Theta_0) = & 
    - \frac{\av{k}\av{k^3}}{\av{k^2}^2} \Theta_0^2
    \frac{\partial}{\partial \Theta_0} S(t|\Theta_0) \\
    & + \frac{\av{k^3}}{\av{k} \av{k^2}} \frac{\Theta_0}{N} 
    \frac{\partial^2}{\partial \Theta_0^2}  S(t|\Theta_0) \, ,
    \end{split}
    \label{eq:Surv_Theta0}
\end{equation}
with the boundary conditions
\begin{equation}
    S(t|\Theta_0 = 0) = 0 \,\,\, \text{and} \,\,\,\left. \frac{\partial}{\partial \Theta_0}  S(t|\Theta_0)\right|_{\Theta_0=1} = 0 \, ,
    \label{eq:surv_bounds}
\end{equation}
expressing the absorbing nature of the barrier in $\Theta=0$ and the reflecting
nature of the barrier in $\Theta=1$. The initial condition on $\Theta$
explicitly depends on the system size as 
$\Theta_0 = k_0 / (N\av{k})$. 
We are interested in deriving the decay of $S(t)$ over time and this requires the thermodynamic limit
to be taken, so we need an initial condition that does not depend on the 
system size~\cite{boguna2009langevin}.
We change variable from $\Theta_0 $ to $k_0=\Theta_0 N\av{k}$. In this way, if we fix
the initial condition to be such that only a single node with degree $k_0=\av{k}$
is infected, then $\Theta_0 = 1/N$.
We obtain
\begin{equation}
     \frac{\partial}{\partial t} S(t|k_0) = 
     -\frac{\av{k^3}}{\av{k^2}^2} \frac{k_0^2}{N} 
     \frac{\partial}{\partial k_0} S(t|k_0) + 
     \frac{\av{k^3}}{\av{k^2}} k_0 
     \frac{\partial^2}{\partial k_0^2}  S(t|k_0) \, .
    \label{eq:Surv_k0}
\end{equation}
where now the thermodynamic limit $N \to \infty$ can be taken. 
The coefficient of the drift term scales as 
\begin{equation}
    \frac{\av{k^3}}{\av{k^2}^2}N^{-1} \sim 
        k_{max}^{\gamma - (2+\omega)}
\end{equation}
if $2 < \gamma \leq 3$ while it scales as $k_{max}^{-\omega}$
if  $4 < \gamma$. Analogously, 
the diffusion coefficient scales as 
\begin{equation}
    \frac{\av{k^3}}{\av{k^2}} \sim 
        k_{max}
    \label{eq:diff_scaling}
\end{equation}
if $2 < \gamma \leq 3$ and it converges to a constant for $\gamma > 4$.


For $\gamma>4$ we have a ``standard" (in the sense of Ref.~\cite{di2017simple}) 
diffusive scenario where the drift 
term tends to zero (for any $\omega > 0$) and the diffusion coefficient converges
to a constant.
For $2 \leq \gamma < 3$ the diffusion 
coefficient diverges and 
the drift term again tends to zero unless $\gamma > 2+\omega$, which is 
impossible if $\omega \geq \gamma-1$. This condition is always met in our numerical
simulations when $2 < \gamma < 3$.
It follows that for large $N$
the dynamics is dominated by the diffusion term, for both homogeneous and heterogeneous 
networks, in complete
analogy with the CP~\cite{boguna2009langevin}. In the limit of large but finite
$N$ the drift term can be neglected and the solution to the resulting equation is
Eq.~(\ref{eq:St}).

The finite-size cutoff of $S(t)$ can be estimated from the average duration
of avalanches~\cite{boguna2009langevin}. For it to be computed, we first need to compute
$T_1(k_0) = \int_0^{\infty} S(t|k_0) dt$.
Integrating in time Eq.~(\ref{eq:Surv_k0}) we obtain
\begin{equation}
    \frac{d^2T_1(k_0)}{dk_0^2} - \frac{k_0}{N \av{k^2}} \frac{dT_1(k_0)}{dk_0} = 
    - \frac{\av{k^2}}{\av{k^3}k_0} \, ,
    \label{eq:T1}
\end{equation}
with boundary conditions, obtained integrating Eqs.~(\ref{eq:surv_bounds}), given by
$T_1(0)=0$ and $T'_1(k_{tot})=0$, where 
$k_{tot}=N\av{k}/2$ is the total number of edges.
The solution is given by
\begin{equation}
    \begin{split}
    & T_1(k_0) = \\ & \sqrt{2N\av{k^2}} 
    \int_0^{k_0/\sqrt{2N\av{k^2}}} du \, e^{u^2} 
    \int_u^{\sqrt{N\av{k}^2/8\av{k^2}}} \frac{\av{k^2}}{\av{k^3}}
    \frac{dv}{v} \, e^{-v^2} \, ,
    \end{split}
\end{equation}
as can be verified by direct substitution.

Defining $T_2(k_0) = 2 \int_0^{\infty}t S(t|k_0)$ and integrating again
Eq.~(\ref{eq:Surv_k0}) we get
\begin{equation}
    \frac{d^2 T_2(k_0)}{d k_0^2} - \frac{k_0}{N \av{k^2}}\frac{dT_2(k_0)}{dk_0} = 
    - \frac{2}{k_0} \frac{\av{k^2}}{\av{k^3}} T_1(k_0) \, .
    \label{eq:T_2}
\end{equation}
The solution is given by
\begin{equation}
    T_2(k_0) = 2 N \av{k^2} \int_0^{k_0/\sqrt{2 N \av{k^2}}} du e^{u^2} 
    \int_u^{\sqrt{N\av{k}^2/8\av{k^2}}} G(t) e^{-t^2} dt \, ,
    \label{eq:T2_sol}
\end{equation}
with
\begin{equation}
    G(t) = 2 \left( \frac{\av{k^2}}{\av{k^3}} \right)^2 t^{-1} \int_0^t e^{u^2} du
    \int_u^{\sqrt{N\av{k}^2/8\av{k^2}}} \frac{dv}{v} e^{-v^2} \, .
\end{equation}
For large $N$ we can approximate the upper limit of the outer integral
in Eq.~(\ref{eq:T2_sol}) with 0 and noting that $\sqrt{N\av{k}^2/8\av{k^2}}$ 
diverges we get
\begin{equation}
    T_2(k_0) \approx k_0\sqrt{2 N \av{k^2}}
    \int_0^{\infty} G(t) e^{-t^2} dt \, ,
\end{equation}
so that the size-dependent cutoff time $t_{\times}(N)$ scales according to 
Eq.~(\ref{eq:tc_homog}) if $\av{k^3}$ is asymptotically finite ($\gamma > 4$),
while it scales as Eq.~(\ref{eq:tc}) if $2 < \gamma \leq 3$.

\section{Determination of the critical point in quenched networks}
\label{app:suscept}

\begin{figure}
\begin{center}
\includegraphics[width=0.45\textwidth]{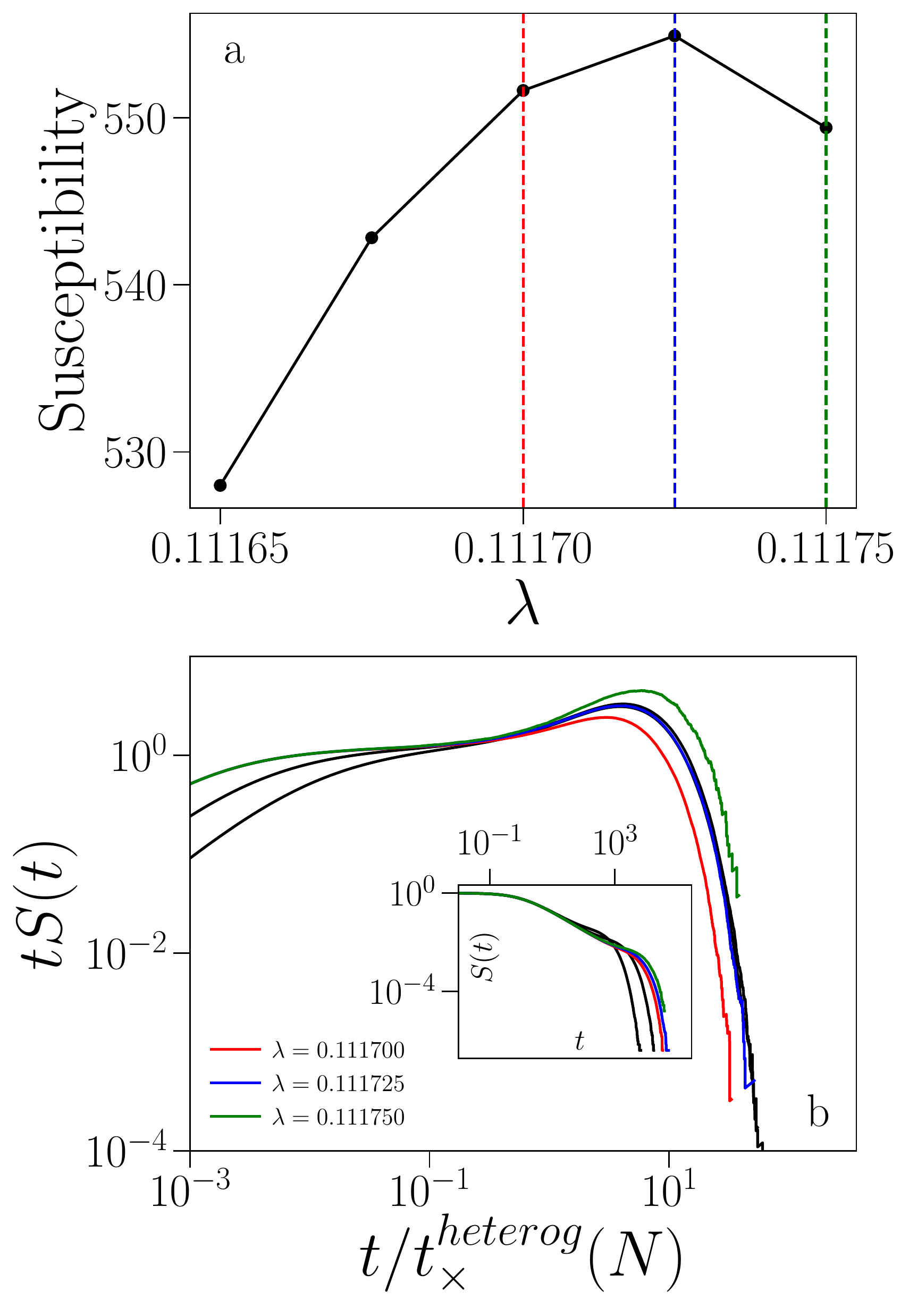} 
\end{center}
\caption{Critical point determination and its importance. (a) The susceptibility
for the random regular graph with $N=10^6$ and $k=10$ is computed for different
values of $\lambda$, with a resolution $2.5 \cdot 10^{-5}$. 
(b) Rescaled survival probability for the random regular graph with $k=10$. 
The abscissa is rescaled as $t/t_c(N)$ and the ordinate is rescaled as $t S(t)$.
The inset shows the same data but the axes are not rescaled. The two black
curves represent $N=10^4$ and $N=10^5$. The red, blue and green lines represent data
for $N=10^6$ obtained using the three values of the $\lambda$ marked by 
the dashed lines of the same color in panel (a).}
\label{fig:suscept}
\end{figure}

Figure~\ref{fig:suscept} shows the importance of a precise estimation of the critical 
point on quenched networks. Figure~\ref{fig:suscept}a shows the susceptibility 
obtained on a random graph with 
$N=10^6$ as $\lambda$ is varied. The critical point is determined as the value
of the spreading rate that maximizes the susceptibility, marked by the blue
dashed line. Figure~\ref{fig:suscept}b shows three curves obtained using three different
values of $\lambda$, the critical one and the two closest values for which we studied
the susceptibility (resolution set to $2.5 \cdot 10^{-5}$). 
Despite the curves look extremely similar (inset), the data collapse is strongly
impacted by $\lambda$. Figure~\ref{fig:suscept}b shows this by slightly increasing or decreasing $\lambda$ with respect to to the 
critical value. 




\begin{thebibliography}{10}
\expandafter\ifx\csname url\endcsname\relax
  \def\url#1{\texttt{#1}}\fi
\expandafter\ifx\csname urlprefix\endcsname\relax\def\urlprefix{URL }\fi
\providecommand{\bibinfo}[2]{#2}
\providecommand{\eprint}[2][]{\url{#2}}

\section*{References}
\makeatletter
\providecommand \@ifxundefined [1]{%
 \@ifx{#1\undefined}
}%
\providecommand \@ifnum [1]{%
 \ifnum #1\expandafter \@firstoftwo
 \else \expandafter \@secondoftwo
 \fi
}%
\providecommand \@ifx [1]{%
 \ifx #1\expandafter \@firstoftwo
 \else \expandafter \@secondoftwo
 \fi
}%
\providecommand \natexlab [1]{#1}%
\providecommand \enquote  [1]{``#1''}%
\providecommand \bibnamefont  [1]{#1}%
\providecommand \bibfnamefont [1]{#1}%
\providecommand \citenamefont [1]{#1}%
\providecommand \href@noop [0]{\@secondoftwo}%
\providecommand \href [0]{\begingroup \@sanitize@url \@href}%
\providecommand \@href[1]{\@@startlink{#1}\@@href}%
\providecommand \@@href[1]{\endgroup#1\@@endlink}%
\providecommand \@sanitize@url [0]{\catcode `\\12\catcode `\$12\catcode
  `\&12\catcode `\#12\catcode `\^12\catcode `\_12\catcode `\%12\relax}%
\providecommand \@@startlink[1]{}%
\providecommand \@@endlink[0]{}%
\providecommand \url  [0]{\begingroup\@sanitize@url \@url }%
\providecommand \@url [1]{\endgroup\@href {#1}{\urlprefix }}%
\providecommand \urlprefix  [0]{URL }%
\providecommand \Eprint [0]{\href }%
\providecommand \doibase [0]{https://doi.org/}%
\providecommand \selectlanguage [0]{\@gobble}%
\providecommand \bibinfo  [0]{\@secondoftwo}%
\providecommand \bibfield  [0]{\@secondoftwo}%
\providecommand \translation [1]{[#1]}%
\providecommand \BibitemOpen [0]{}%
\providecommand \bibitemStop [0]{}%
\providecommand \bibitemNoStop [0]{.\EOS\space}%
\providecommand \EOS [0]{\spacefactor3000\relax}%
\providecommand \BibitemShut  [1]{\csname bibitem#1\endcsname}%
\let\auto@bib@innerbib\@empty
\bibitem [{\citenamefont {Beggs}\ and\ \citenamefont
  {Plenz}(2003)}]{beggs2003neuronal}%
  \BibitemOpen
  \bibfield  {author} {\bibinfo {author} {\bibfnamefont {J.}~\bibnamefont
  {Beggs}}\ and\ \bibinfo {author} {\bibfnamefont {D.}~\bibnamefont {Plenz}},\
  }\bibfield  {title} {\bibinfo {title} {Neuronal avalanches in neocortical
  circuits},\ }\href@noop {} {\bibfield  {journal} {\bibinfo  {journal} {J.
  Neurosci.}\ }\textbf {\bibinfo {volume} {23}},\ \bibinfo {pages} {11167}
  (\bibinfo {year} {2003})}\BibitemShut {NoStop}%
\bibitem [{\citenamefont {Wang}\ and\ \citenamefont
  {Dai}(2013)}]{wang2013self}%
  \BibitemOpen
  \bibfield  {author} {\bibinfo {author} {\bibfnamefont {F.}~\bibnamefont
  {Wang}}\ and\ \bibinfo {author} {\bibfnamefont {Z.}~\bibnamefont {Dai}},\
  }\bibfield  {title} {\bibinfo {title} {Self-organized criticality in x-ray
  flares of gamma-ray-burst afterglows},\ }\href@noop {} {\bibfield  {journal}
  {\bibinfo  {journal} {Nat. Phys.}\ }\textbf {\bibinfo {volume} {9}},\
  \bibinfo {pages} {465} (\bibinfo {year} {2013})}\BibitemShut {NoStop}%
\bibitem [{\citenamefont {Bak}\ \emph {et~al.}(2002)\citenamefont {Bak},
  \citenamefont {Christensen}, \citenamefont {Danon},\ and\ \citenamefont
  {Scanlon}}]{bak2002unified}%
  \BibitemOpen
  \bibfield  {author} {\bibinfo {author} {\bibfnamefont {P.}~\bibnamefont
  {Bak}}, \bibinfo {author} {\bibfnamefont {K.}~\bibnamefont {Christensen}},
  \bibinfo {author} {\bibfnamefont {L.}~\bibnamefont {Danon}},\ and\ \bibinfo
  {author} {\bibfnamefont {T.}~\bibnamefont {Scanlon}},\ }\bibfield  {title}
  {\bibinfo {title} {Unified scaling law for earthquakes},\ }\href@noop {}
  {\bibfield  {journal} {\bibinfo  {journal} {Phys. Rev. Lett.}\ }\textbf
  {\bibinfo {volume} {88}} (\bibinfo {year} {2002})}\BibitemShut {NoStop}%
\bibitem [{\citenamefont {Kinney}\ \emph {et~al.}(2005)\citenamefont {Kinney},
  \citenamefont {Crucitti}, \citenamefont {Albert},\ and\ \citenamefont
  {Latora}}]{kinney2005modeling}%
  \BibitemOpen
  \bibfield  {author} {\bibinfo {author} {\bibfnamefont {R.}~\bibnamefont
  {Kinney}}, \bibinfo {author} {\bibfnamefont {P.}~\bibnamefont {Crucitti}},
  \bibinfo {author} {\bibfnamefont {R.}~\bibnamefont {Albert}},\ and\ \bibinfo
  {author} {\bibfnamefont {V.}~\bibnamefont {Latora}},\ }\bibfield  {title}
  {\bibinfo {title} {Modeling cascading failures in the north american power
  grid},\ }\href@noop {} {\bibfield  {journal} {\bibinfo  {journal} {EPJB}\
  }\textbf {\bibinfo {volume} {46}},\ \bibinfo {pages} {101} (\bibinfo {year}
  {2005})}\BibitemShut {NoStop}%
\bibitem [{\citenamefont {Nishi}\ \emph {et~al.}(2016)\citenamefont {Nishi},
  \citenamefont {Takaguchi}, \citenamefont {Oka}, \citenamefont {Maehara},
  \citenamefont {Toyoda}, \citenamefont {Kawarabayashi},\ and\ \citenamefont
  {Masuda}}]{nishi2016reply}%
  \BibitemOpen
  \bibfield  {author} {\bibinfo {author} {\bibfnamefont {R.}~\bibnamefont
  {Nishi}}, \bibinfo {author} {\bibfnamefont {T.}~\bibnamefont {Takaguchi}},
  \bibinfo {author} {\bibfnamefont {K.}~\bibnamefont {Oka}}, \bibinfo {author}
  {\bibfnamefont {T.}~\bibnamefont {Maehara}}, \bibinfo {author} {\bibfnamefont
  {M.}~\bibnamefont {Toyoda}}, \bibinfo {author} {\bibfnamefont
  {K.}~\bibnamefont {Kawarabayashi}},\ and\ \bibinfo {author} {\bibfnamefont
  {N.}~\bibnamefont {Masuda}},\ }\bibfield  {title} {\bibinfo {title} {Reply
  trees in twitter: data analysis and branching process models},\ }\href@noop
  {} {\bibfield  {journal} {\bibinfo  {journal} {SNAM}\ }\textbf {\bibinfo
  {volume} {6}},\ \bibinfo {pages} {26} (\bibinfo {year} {2016})}\BibitemShut
  {NoStop}%
\bibitem [{\citenamefont {Wegrzycki}\ \emph {et~al.}(2017)\citenamefont
  {Wegrzycki}, \citenamefont {Sankowski}, \citenamefont {Pacuk},\ and\
  \citenamefont {Wygocki}}]{wegrzycki2017cascade}%
  \BibitemOpen
  \bibfield  {author} {\bibinfo {author} {\bibfnamefont {K.}~\bibnamefont
  {Wegrzycki}}, \bibinfo {author} {\bibfnamefont {P.}~\bibnamefont
  {Sankowski}}, \bibinfo {author} {\bibfnamefont {A.}~\bibnamefont {Pacuk}},\
  and\ \bibinfo {author} {\bibfnamefont {P.}~\bibnamefont {Wygocki}},\
  }\bibfield  {title} {\bibinfo {title} {Why do cascade sizes follow a
  power-law?},\ }\href@noop {} {\bibfield  {journal} {\bibinfo  {journal}
  {Proceedings of the 26th International Conference on World Wide Web}\ }
  (\bibinfo {year} {2017})}\BibitemShut {NoStop}%
\bibitem [{\citenamefont {Lerman}\ and\ \citenamefont
  {Ghosh}(2010)}]{lerman2010information}%
  \BibitemOpen
  \bibfield  {author} {\bibinfo {author} {\bibfnamefont {K.}~\bibnamefont
  {Lerman}}\ and\ \bibinfo {author} {\bibfnamefont {R.}~\bibnamefont {Ghosh}},\
  }\bibfield  {title} {\bibinfo {title} {Information contagion: An empirical
  study of the spread of news on digg and twitter social networks},\ }in\
  \href@noop {} {\emph {\bibinfo {booktitle} {Fourth International AAAI
  Conference on Weblogs and Social Media}}}\ (\bibinfo {year}
  {2010})\BibitemShut {NoStop}%
\bibitem [{\citenamefont {Kadanoff}\ \emph {et~al.}(1989)\citenamefont
  {Kadanoff}, \citenamefont {Nagel}, \citenamefont {Wu},\ and\ \citenamefont
  {Zhou}}]{Kadanoff1989}%
  \BibitemOpen
  \bibfield  {author} {\bibinfo {author} {\bibfnamefont {L.~P.}\ \bibnamefont
  {Kadanoff}}, \bibinfo {author} {\bibfnamefont {S.~R.}\ \bibnamefont {Nagel}},
  \bibinfo {author} {\bibfnamefont {L.}~\bibnamefont {Wu}},\ and\ \bibinfo
  {author} {\bibfnamefont {S.-m.}\ \bibnamefont {Zhou}},\ }\bibfield  {title}
  {\bibinfo {title} {Scaling and universality in avalanches},\ }\href
  {https://doi.org/10.1103/PhysRevA.39.6524} {\bibfield  {journal} {\bibinfo
  {journal} {Phys. Rev. A}\ }\textbf {\bibinfo {volume} {39}},\ \bibinfo
  {pages} {6524} (\bibinfo {year} {1989})}\BibitemShut {NoStop}%
\bibitem [{\citenamefont {Sethna}\ \emph {et~al.}(2001)\citenamefont {Sethna},
  \citenamefont {Dahmen},\ and\ \citenamefont {Myers}}]{sethna2001crackling}%
  \BibitemOpen
  \bibfield  {author} {\bibinfo {author} {\bibfnamefont {J.~P.}\ \bibnamefont
  {Sethna}}, \bibinfo {author} {\bibfnamefont {K.~A.}\ \bibnamefont {Dahmen}},\
  and\ \bibinfo {author} {\bibfnamefont {C.~R.}\ \bibnamefont {Myers}},\
  }\bibfield  {title} {\bibinfo {title} {Crackling noise},\ }\href@noop {}
  {\bibfield  {journal} {\bibinfo  {journal} {Nature}\ }\textbf {\bibinfo
  {volume} {410}},\ \bibinfo {pages} {242} (\bibinfo {year}
  {2001})}\BibitemShut {NoStop}%
\bibitem [{\citenamefont {Harris}\ \emph {et~al.}(1963)\citenamefont {Harris}
  \emph {et~al.}}]{harris1963theory}%
  \BibitemOpen
  \bibfield  {author} {\bibinfo {author} {\bibfnamefont {T.~E.}\ \bibnamefont
  {Harris}} \emph {et~al.},\ }\href@noop {} {\emph {\bibinfo {title} {The
  theory of branching processes}}},\ Vol.~\bibinfo {volume} {6}\ (\bibinfo
  {publisher} {Springer Berlin},\ \bibinfo {year} {1963})\BibitemShut {NoStop}%
\bibitem [{\citenamefont {Adami}\ and\ \citenamefont {Chu}(2002)}]{Adami2002}%
  \BibitemOpen
  \bibfield  {author} {\bibinfo {author} {\bibfnamefont {C.}~\bibnamefont
  {Adami}}\ and\ \bibinfo {author} {\bibfnamefont {J.}~\bibnamefont {Chu}},\
  }\bibfield  {title} {\bibinfo {title} {Critical and near-critical branching
  processes},\ }\href {https://doi.org/10.1103/PhysRevE.66.011907} {\bibfield
  {journal} {\bibinfo  {journal} {Phys. Rev. E}\ }\textbf {\bibinfo {volume}
  {66}},\ \bibinfo {pages} {011907} (\bibinfo {year} {2002})}\BibitemShut
  {NoStop}%
\bibitem [{\citenamefont {Goh}\ \emph {et~al.}(2003)\citenamefont {Goh},
  \citenamefont {Lee}, \citenamefont {Kahng},\ and\ \citenamefont
  {Kim}}]{Goh2003}%
  \BibitemOpen
  \bibfield  {author} {\bibinfo {author} {\bibfnamefont {K.-I.}\ \bibnamefont
  {Goh}}, \bibinfo {author} {\bibfnamefont {D.-S.}\ \bibnamefont {Lee}},
  \bibinfo {author} {\bibfnamefont {B.}~\bibnamefont {Kahng}},\ and\ \bibinfo
  {author} {\bibfnamefont {D.}~\bibnamefont {Kim}},\ }\bibfield  {title}
  {\bibinfo {title} {Sandpile on scale-free networks},\ }\href@noop {}
  {\bibfield  {journal} {\bibinfo  {journal} {Phys. Rev. Lett.}\ }\textbf
  {\bibinfo {volume} {91}},\ \bibinfo {pages} {148701} (\bibinfo {year}
  {2003})}\BibitemShut {NoStop}%
\bibitem [{\citenamefont {Saichev}\ \emph {et~al.}(2005)\citenamefont
  {Saichev}, \citenamefont {Helmstetter},\ and\ \citenamefont
  {Sornette}}]{Saichev2005}%
  \BibitemOpen
  \bibfield  {author} {\bibinfo {author} {\bibfnamefont {A.}~\bibnamefont
  {Saichev}}, \bibinfo {author} {\bibfnamefont {A.}~\bibnamefont
  {Helmstetter}},\ and\ \bibinfo {author} {\bibfnamefont {D.}~\bibnamefont
  {Sornette}},\ }\bibfield  {title} {\bibinfo {title} {Power-law distributions
  of offspring and generation numbers in branching models of earthquake trig
  gering},\ }\href@noop {} {\bibfield  {journal} {\bibinfo  {journal} {Pure
  Appl. Geophys.}\ }\textbf {\bibinfo {volume} {162}},\ \bibinfo {pages} {1113}
  (\bibinfo {year} {2005})}\BibitemShut {NoStop}%
\bibitem [{\citenamefont {Gleeson}\ \emph {et~al.}(2014)\citenamefont
  {Gleeson}, \citenamefont {Ward}, \citenamefont {O'Sullivan},\ and\
  \citenamefont {Lee}}]{gleeson2014competition}%
  \BibitemOpen
  \bibfield  {author} {\bibinfo {author} {\bibfnamefont {J.}~\bibnamefont
  {Gleeson}}, \bibinfo {author} {\bibfnamefont {J.}~\bibnamefont {Ward}},
  \bibinfo {author} {\bibfnamefont {K.}~\bibnamefont {O'Sullivan}},\ and\
  \bibinfo {author} {\bibfnamefont {W.}~\bibnamefont {Lee}},\ }\bibfield
  {title} {\bibinfo {title} {Competition-induced criticality in a model of meme
  popularity.},\ }\href@noop {} {\bibfield  {journal} {\bibinfo  {journal}
  {Phys. Rev. Lett.}\ }\textbf {\bibinfo {volume} {112}},\ \bibinfo {pages}
  {048701} (\bibinfo {year} {2014})}\BibitemShut {NoStop}%
\bibitem [{\citenamefont {di~Santo}\ \emph {et~al.}(2017)\citenamefont
  {di~Santo}, \citenamefont {Villegas}, \citenamefont {Burioni},\ and\
  \citenamefont {Mu\~noz}}]{di2017simple}%
  \BibitemOpen
  \bibfield  {author} {\bibinfo {author} {\bibfnamefont {S.}~\bibnamefont
  {di~Santo}}, \bibinfo {author} {\bibfnamefont {P.}~\bibnamefont {Villegas}},
  \bibinfo {author} {\bibfnamefont {R.}~\bibnamefont {Burioni}},\ and\ \bibinfo
  {author} {\bibfnamefont {M.~A.}\ \bibnamefont {Mu\~noz}},\ }\bibfield
  {title} {\bibinfo {title} {Simple unified view of branching process
  statistics: Random walks in balanced logarithmic potentials},\ }\href
  {https://doi.org/10.1103/PhysRevE.95.032115} {\bibfield  {journal} {\bibinfo
  {journal} {Phys. Rev. E}\ }\textbf {\bibinfo {volume} {95}},\ \bibinfo
  {pages} {032115} (\bibinfo {year} {2017})}\BibitemShut {NoStop}%
\bibitem [{\citenamefont {Zapperi}\ \emph {et~al.}(1995)\citenamefont
  {Zapperi}, \citenamefont {Lauritsen},\ and\ \citenamefont
  {Stanley}}]{Zapperi1995}%
  \BibitemOpen
  \bibfield  {author} {\bibinfo {author} {\bibfnamefont {S.}~\bibnamefont
  {Zapperi}}, \bibinfo {author} {\bibfnamefont {K.~B.}\ \bibnamefont
  {Lauritsen}},\ and\ \bibinfo {author} {\bibfnamefont {H.~E.}\ \bibnamefont
  {Stanley}},\ }\bibfield  {title} {\bibinfo {title} {Self-organized branching
  processes: Mean-field theory for avalanches},\ }\href
  {https://doi.org/10.1103/PhysRevLett.75.4071} {\bibfield  {journal} {\bibinfo
   {journal} {Phys. Rev. Lett.}\ }\textbf {\bibinfo {volume} {75}},\ \bibinfo
  {pages} {4071} (\bibinfo {year} {1995})}\BibitemShut {NoStop}%
\bibitem [{\citenamefont {Gleeson}\ and\ \citenamefont
  {Durrett}(2017)}]{gleeson2017temporal}%
  \BibitemOpen
  \bibfield  {author} {\bibinfo {author} {\bibfnamefont {J.~P.}\ \bibnamefont
  {Gleeson}}\ and\ \bibinfo {author} {\bibfnamefont {R.}~\bibnamefont
  {Durrett}},\ }\bibfield  {title} {\bibinfo {title} {Temporal profiles of
  avalanches on networks},\ }\href@noop {} {\bibfield  {journal} {\bibinfo
  {journal} {Nature communications}\ }\textbf {\bibinfo {volume} {8}},\
  \bibinfo {pages} {1} (\bibinfo {year} {2017})}\BibitemShut {NoStop}%
\bibitem [{\citenamefont {Matin}\ \emph {et~al.}(2021)\citenamefont {Matin},
  \citenamefont {Tenzin},\ and\ \citenamefont {Klein}}]{matin2021scaling}%
  \BibitemOpen
  \bibfield  {author} {\bibinfo {author} {\bibfnamefont {S.}~\bibnamefont
  {Matin}}, \bibinfo {author} {\bibfnamefont {T.}~\bibnamefont {Tenzin}},\ and\
  \bibinfo {author} {\bibfnamefont {W.}~\bibnamefont {Klein}},\ }\bibfield
  {title} {\bibinfo {title} {Scaling of causal neural avalanches in a neutral
  model},\ }\href {https://doi.org/10.1103/PhysRevResearch.3.013107} {\bibfield
   {journal} {\bibinfo  {journal} {Phys. Rev. Research}\ }\textbf {\bibinfo
  {volume} {3}},\ \bibinfo {pages} {013107} (\bibinfo {year}
  {2021})}\BibitemShut {NoStop}%
\bibitem [{\citenamefont {Radicchi}\ \emph {et~al.}(2020)\citenamefont
  {Radicchi}, \citenamefont {Castellano}, \citenamefont {Flammini},
  \citenamefont {Mu\~noz},\ and\ \citenamefont
  {Notarmuzi}}]{radicchi2020classes}%
  \BibitemOpen
  \bibfield  {author} {\bibinfo {author} {\bibfnamefont {F.}~\bibnamefont
  {Radicchi}}, \bibinfo {author} {\bibfnamefont {C.}~\bibnamefont
  {Castellano}}, \bibinfo {author} {\bibfnamefont {A.}~\bibnamefont
  {Flammini}}, \bibinfo {author} {\bibfnamefont {M.~A.}\ \bibnamefont
  {Mu\~noz}},\ and\ \bibinfo {author} {\bibfnamefont {D.}~\bibnamefont
  {Notarmuzi}},\ }\bibfield  {title} {\bibinfo {title} {Classes of critical
  avalanche dynamics in complex networks},\ }\href
  {https://doi.org/10.1103/PhysRevResearch.2.033171} {\bibfield  {journal}
  {\bibinfo  {journal} {Phys. Rev. Research}\ }\textbf {\bibinfo {volume}
  {2}},\ \bibinfo {pages} {033171} (\bibinfo {year} {2020})}\BibitemShut
  {NoStop}%
\bibitem [{\citenamefont {Larremore}\ \emph {et~al.}(2012)\citenamefont
  {Larremore}, \citenamefont {Carpenter}, \citenamefont {Ott},\ and\
  \citenamefont {Restrepo}}]{larremore2012statistical}%
  \BibitemOpen
  \bibfield  {author} {\bibinfo {author} {\bibfnamefont {D.~B.}\ \bibnamefont
  {Larremore}}, \bibinfo {author} {\bibfnamefont {M.~Y.}\ \bibnamefont
  {Carpenter}}, \bibinfo {author} {\bibfnamefont {E.}~\bibnamefont {Ott}},\
  and\ \bibinfo {author} {\bibfnamefont {J.~G.}\ \bibnamefont {Restrepo}},\
  }\bibfield  {title} {\bibinfo {title} {Statistical properties of avalanches
  in networks},\ }\href {https://doi.org/10.1103/PhysRevE.85.066131} {\bibfield
   {journal} {\bibinfo  {journal} {Phys. Rev. E}\ }\textbf {\bibinfo {volume}
  {85}},\ \bibinfo {pages} {066131} (\bibinfo {year} {2012})}\BibitemShut
  {NoStop}%
\bibitem [{\citenamefont {Pastor-Satorras}\ \emph {et~al.}(2015)\citenamefont
  {Pastor-Satorras}, \citenamefont {Castellano}, \citenamefont {Van~Mieghem},\
  and\ \citenamefont {Vespignani}}]{PastorSatorras2015}%
  \BibitemOpen
  \bibfield  {author} {\bibinfo {author} {\bibfnamefont {R.}~\bibnamefont
  {Pastor-Satorras}}, \bibinfo {author} {\bibfnamefont {C.}~\bibnamefont
  {Castellano}}, \bibinfo {author} {\bibfnamefont {P.}~\bibnamefont
  {Van~Mieghem}},\ and\ \bibinfo {author} {\bibfnamefont {A.}~\bibnamefont
  {Vespignani}},\ }\bibfield  {title} {\bibinfo {title} {Epidemic processes in
  complex networks},\ }\href {https://doi.org/10.1103/RevModPhys.87.925}
  {\bibfield  {journal} {\bibinfo  {journal} {Rev. Mod. Phys.}\ }\textbf
  {\bibinfo {volume} {87}},\ \bibinfo {pages} {925} (\bibinfo {year}
  {2015})}\BibitemShut {NoStop}%
\bibitem [{\citenamefont {Pastor-Satorras}\ and\ \citenamefont
  {Vespignani}(2001)}]{Pastor2001epidemic}%
  \BibitemOpen
  \bibfield  {author} {\bibinfo {author} {\bibfnamefont {R.}~\bibnamefont
  {Pastor-Satorras}}\ and\ \bibinfo {author} {\bibfnamefont {A.}~\bibnamefont
  {Vespignani}},\ }\bibfield  {title} {\bibinfo {title} {Epidemic spreading in
  scale-free networks},\ }\href@noop {} {\bibfield  {journal} {\bibinfo
  {journal} {Phys. Rev. Lett.}\ }\textbf {\bibinfo {volume} {86}},\ \bibinfo
  {pages} {3200} (\bibinfo {year} {2001})}\BibitemShut {NoStop}%
\bibitem [{\citenamefont {Castellano}\ and\ \citenamefont
  {Pastor-Satorras}(2010)}]{castellano2010thresholds}%
  \BibitemOpen
  \bibfield  {author} {\bibinfo {author} {\bibfnamefont {C.}~\bibnamefont
  {Castellano}}\ and\ \bibinfo {author} {\bibfnamefont {R.}~\bibnamefont
  {Pastor-Satorras}},\ }\bibfield  {title} {\bibinfo {title} {Thresholds for
  epidemic spreading in networks},\ }\href@noop {} {\bibfield  {journal}
  {\bibinfo  {journal} {Physical review letters}\ }\textbf {\bibinfo {volume}
  {105}},\ \bibinfo {pages} {218701} (\bibinfo {year} {2010})}\BibitemShut
  {NoStop}%
\bibitem [{\citenamefont {Bogu\~n\'a}\ \emph {et~al.}(2013)\citenamefont
  {Bogu\~n\'a}, \citenamefont {Castellano},\ and\ \citenamefont
  {Pastor-Satorras}}]{Boguna2013}%
  \BibitemOpen
  \bibfield  {author} {\bibinfo {author} {\bibfnamefont {M.}~\bibnamefont
  {Bogu\~n\'a}}, \bibinfo {author} {\bibfnamefont {C.}~\bibnamefont
  {Castellano}},\ and\ \bibinfo {author} {\bibfnamefont {R.}~\bibnamefont
  {Pastor-Satorras}},\ }\bibfield  {title} {\bibinfo {title} {Nature of the
  epidemic threshold for the susceptible-infected-susceptible dynamics in
  networks},\ }\href {https://doi.org/10.1103/PhysRevLett.111.068701}
  {\bibfield  {journal} {\bibinfo  {journal} {Phys. Rev. Lett.}\ }\textbf
  {\bibinfo {volume} {111}},\ \bibinfo {pages} {068701} (\bibinfo {year}
  {2013})}\BibitemShut {NoStop}%
\bibitem [{\citenamefont {Castellano}\ and\ \citenamefont
  {Pastor-Satorras}(2012)}]{castellano2012competing}%
  \BibitemOpen
  \bibfield  {author} {\bibinfo {author} {\bibfnamefont {C.}~\bibnamefont
  {Castellano}}\ and\ \bibinfo {author} {\bibfnamefont {R.}~\bibnamefont
  {Pastor-Satorras}},\ }\bibfield  {title} {\bibinfo {title} {{Competing
  activation mechanisms in epidemics on networks}},\ }\href
  {https://doi.org/10.1038/srep00371} {\bibfield  {journal} {\bibinfo
  {journal} {Scientific Reports}\ }\textbf {\bibinfo {volume} {2}},\ \bibinfo
  {pages} {00371} (\bibinfo {year} {2012})}\BibitemShut {NoStop}%
\bibitem [{\citenamefont {Castellano}\ and\ \citenamefont
  {Pastor-Satorras}(2020)}]{Castellano2020}%
  \BibitemOpen
  \bibfield  {author} {\bibinfo {author} {\bibfnamefont {C.}~\bibnamefont
  {Castellano}}\ and\ \bibinfo {author} {\bibfnamefont {R.}~\bibnamefont
  {Pastor-Satorras}},\ }\bibfield  {title} {\bibinfo {title} {Cumulative
  merging percolation and the epidemic transition of the
  susceptible-infected-susceptible model in networks},\ }\href
  {https://doi.org/10.1103/PhysRevX.10.011070} {\bibfield  {journal} {\bibinfo
  {journal} {Phys. Rev. X}\ }\textbf {\bibinfo {volume} {10}},\ \bibinfo
  {pages} {011070} (\bibinfo {year} {2020})}\BibitemShut {NoStop}%
\bibitem [{\citenamefont {Bogu\~n\'a}\ \emph {et~al.}(2009)\citenamefont
  {Bogu\~n\'a}, \citenamefont {Castellano},\ and\ \citenamefont
  {Pastor-Satorras}}]{boguna2009langevin}%
  \BibitemOpen
  \bibfield  {author} {\bibinfo {author} {\bibfnamefont {M.}~\bibnamefont
  {Bogu\~n\'a}}, \bibinfo {author} {\bibfnamefont {C.}~\bibnamefont
  {Castellano}},\ and\ \bibinfo {author} {\bibfnamefont {R.}~\bibnamefont
  {Pastor-Satorras}},\ }\bibfield  {title} {\bibinfo {title} {Langevin approach
  for the dynamics of the contact process on annealed scale-free networks},\
  }\href {https://doi.org/10.1103/PhysRevE.79.036110} {\bibfield  {journal}
  {\bibinfo  {journal} {Phys. Rev. E}\ }\textbf {\bibinfo {volume} {79}},\
  \bibinfo {pages} {036110} (\bibinfo {year} {2009})}\BibitemShut {NoStop}%
\bibitem [{\citenamefont {Molloy}\ and\ \citenamefont
  {Reed}(1995)}]{molloy1995critical}%
  \BibitemOpen
  \bibfield  {author} {\bibinfo {author} {\bibfnamefont {M.}~\bibnamefont
  {Molloy}}\ and\ \bibinfo {author} {\bibfnamefont {B.}~\bibnamefont {Reed}},\
  }\bibfield  {title} {\bibinfo {title} {A critical point for random graphs
  with a given degree sequence},\ }\href@noop {} {\bibfield  {journal}
  {\bibinfo  {journal} {Random structures \& algorithms}\ }\textbf {\bibinfo
  {volume} {6}},\ \bibinfo {pages} {161} (\bibinfo {year} {1995})}\BibitemShut
  {NoStop}%
\bibitem [{\citenamefont {Dorogovtsev}\ \emph {et~al.}(2008)\citenamefont
  {Dorogovtsev}, \citenamefont {Goltsev},\ and\ \citenamefont
  {Mendes}}]{Dorogovtsev2008}%
  \BibitemOpen
  \bibfield  {author} {\bibinfo {author} {\bibfnamefont {S.~N.}\ \bibnamefont
  {Dorogovtsev}}, \bibinfo {author} {\bibfnamefont {A.~V.}\ \bibnamefont
  {Goltsev}},\ and\ \bibinfo {author} {\bibfnamefont {J.~F.~F.}\ \bibnamefont
  {Mendes}},\ }\bibfield  {title} {\bibinfo {title} {Critical phenomena in
  complex networks},\ }\href {https://doi.org/10.1103/RevModPhys.80.1275}
  {\bibfield  {journal} {\bibinfo  {journal} {Rev. Mod. Phys.}\ }\textbf
  {\bibinfo {volume} {80}},\ \bibinfo {pages} {1275} (\bibinfo {year}
  {2008})}\BibitemShut {NoStop}%
\bibitem [{\citenamefont {Catanzaro}\ \emph {et~al.}(2005)\citenamefont
  {Catanzaro}, \citenamefont {Bogu\~n\'a},\ and\ \citenamefont
  {Pastor-Satorras}}]{Catanzaro2005}%
  \BibitemOpen
  \bibfield  {author} {\bibinfo {author} {\bibfnamefont {M.}~\bibnamefont
  {Catanzaro}}, \bibinfo {author} {\bibfnamefont {M.}~\bibnamefont
  {Bogu\~n\'a}},\ and\ \bibinfo {author} {\bibfnamefont {R.}~\bibnamefont
  {Pastor-Satorras}},\ }\bibfield  {title} {\bibinfo {title} {Generation of
  uncorrelated random scale-free networks},\ }\href
  {https://doi.org/10.1103/PhysRevE.71.027103} {\bibfield  {journal} {\bibinfo
  {journal} {Phys. Rev. E}\ }\textbf {\bibinfo {volume} {71}},\ \bibinfo
  {pages} {027103} (\bibinfo {year} {2005})}\BibitemShut {NoStop}%
\bibitem [{\citenamefont {Garcia-Millan}\ \emph {et~al.}(2018)\citenamefont
  {Garcia-Millan}, \citenamefont {Pausch}, \citenamefont {Walter},\ and\
  \citenamefont {Pruessner}}]{garciamillan2018field}%
  \BibitemOpen
  \bibfield  {author} {\bibinfo {author} {\bibfnamefont {R.}~\bibnamefont
  {Garcia-Millan}}, \bibinfo {author} {\bibfnamefont {J.}~\bibnamefont
  {Pausch}}, \bibinfo {author} {\bibfnamefont {B.}~\bibnamefont {Walter}},\
  and\ \bibinfo {author} {\bibfnamefont {G.}~\bibnamefont {Pruessner}},\
  }\bibfield  {title} {\bibinfo {title} {Field-theoretic approach to the
  universality of branching processes},\ }\href
  {https://doi.org/10.1103/PhysRevE.98.062107} {\bibfield  {journal} {\bibinfo
  {journal} {Phys. Rev. E}\ }\textbf {\bibinfo {volume} {98}},\ \bibinfo
  {pages} {062107} (\bibinfo {year} {2018})}\BibitemShut {NoStop}%
\bibitem [{\citenamefont {Henkel}\ \emph {et~al.}(2008)\citenamefont {Henkel},
  \citenamefont {Hinrichsen}, \citenamefont {L{\"u}beck},\ and\ \citenamefont
  {Pleimling}}]{henkel2008non}%
  \BibitemOpen
  \bibfield  {author} {\bibinfo {author} {\bibfnamefont {M.}~\bibnamefont
  {Henkel}}, \bibinfo {author} {\bibfnamefont {H.}~\bibnamefont {Hinrichsen}},
  \bibinfo {author} {\bibfnamefont {S.}~\bibnamefont {L{\"u}beck}},\ and\
  \bibinfo {author} {\bibfnamefont {M.}~\bibnamefont {Pleimling}},\ }\href@noop
  {} {\emph {\bibinfo {title} {Non-equilibrium phase transitions}}},\
  Vol.~\bibinfo {volume} {1}\ (\bibinfo  {publisher} {Springer},\ \bibinfo
  {year} {2008})\BibitemShut {NoStop}%
\bibitem [{\citenamefont {Pastor-Satorras}\ and\ \citenamefont
  {Castellano}(2016)}]{pastor2016distinct}%
  \BibitemOpen
  \bibfield  {author} {\bibinfo {author} {\bibfnamefont {R.}~\bibnamefont
  {Pastor-Satorras}}\ and\ \bibinfo {author} {\bibfnamefont {C.}~\bibnamefont
  {Castellano}},\ }\bibfield  {title} {\bibinfo {title} {{Distinct types of
  eigenvector localization in networks}},\ }\href
  {https://doi.org/10.1038/srep18847} {\bibfield  {journal} {\bibinfo
  {journal} {Scientific Reports}\ }\textbf {\bibinfo {volume} {6}},\ \bibinfo
  {pages} {18847} (\bibinfo {year} {2016})}\BibitemShut {NoStop}%
\bibitem [{\citenamefont {Gardiner}\ \emph {et~al.}(1985)\citenamefont
  {Gardiner} \emph {et~al.}}]{gardiner1985handbook}%
  \BibitemOpen
  \bibfield  {author} {\bibinfo {author} {\bibfnamefont {C.~W.}\ \bibnamefont
  {Gardiner}} \emph {et~al.},\ }\href@noop {} {\emph {\bibinfo {title}
  {Handbook of stochastic methods}}},\ Vol.~\bibinfo {volume} {3}\ (\bibinfo
  {publisher} {springer Berlin},\ \bibinfo {year} {1985})\BibitemShut {NoStop}%
\bibitem [{\citenamefont {Ferreira}\ \emph {et~al.}(2012)\citenamefont
  {Ferreira}, \citenamefont {Castellano},\ and\ \citenamefont
  {Pastor-Satorras}}]{Ferreira2012}%
  \BibitemOpen
  \bibfield  {author} {\bibinfo {author} {\bibfnamefont {S.~C.}\ \bibnamefont
  {Ferreira}}, \bibinfo {author} {\bibfnamefont {C.}~\bibnamefont
  {Castellano}},\ and\ \bibinfo {author} {\bibfnamefont {R.}~\bibnamefont
  {Pastor-Satorras}},\ }\bibfield  {title} {\bibinfo {title} {Epidemic
  thresholds of the susceptible-infected-susceptible model on networks: A
  comparison of numerical and theoretical results},\ }\href
  {https://doi.org/10.1103/PhysRevE.86.041125} {\bibfield  {journal} {\bibinfo
  {journal} {Phys. Rev. E}\ }\textbf {\bibinfo {volume} {86}},\ \bibinfo
  {pages} {041125} (\bibinfo {year} {2012})}\BibitemShut {NoStop}%
\bibitem [{\citenamefont {Dorogovtsev}\ \emph {et~al.}(2006)\citenamefont
  {Dorogovtsev}, \citenamefont {Goltsev},\ and\ \citenamefont
  {Mendes}}]{dorogovtsev2006k}%
  \BibitemOpen
  \bibfield  {author} {\bibinfo {author} {\bibfnamefont {S.~N.}\ \bibnamefont
  {Dorogovtsev}}, \bibinfo {author} {\bibfnamefont {A.~V.}\ \bibnamefont
  {Goltsev}},\ and\ \bibinfo {author} {\bibfnamefont {J.~F.~F.}\ \bibnamefont
  {Mendes}},\ }\bibfield  {title} {\bibinfo {title} {$k$-core organization of
  complex networks},\ }\href {https://doi.org/10.1103/PhysRevLett.96.040601}
  {\bibfield  {journal} {\bibinfo  {journal} {Phys. Rev. Lett.}\ }\textbf
  {\bibinfo {volume} {96}},\ \bibinfo {pages} {040601} (\bibinfo {year}
  {2006})}\BibitemShut {NoStop}%
\end{thebibliography}


%

\end{document}